\documentclass[aps,prl,twocolumn,footinbib,superscriptaddress]{revtex4-1}
\usepackage{graphicx}
\usepackage{amsmath,amsthm,amssymb,mathtools}
\usepackage{color}

\usepackage[none]{hyphenat}

\def\blue#1{{\color{blue}#1}}
\usepackage{tcolorbox}
\usepackage{cancel}
\usepackage{braket}
\def\shrinkage{2.1mu}
\def\vecsign{\mathchar"017E}
\def\dvecsign{\smash{\stackon[-1.95pt]{\mkern-\shrinkage\vecsign}{\rotatebox{180}{$\mkern-\shrinkage\vecsign$}}}}
\def\dvec#1{\def\useanchorwidth{T}\stackon[-4.2pt]{#1}{\,\dvecsign}}
\usepackage{stackengine}
\stackMath
\usepackage[pdftex,colorlinks=true,linkcolor=Cerulean,citecolor=Cerulean,urlcolor=Cerulean]{hyperref}
\hypersetup{linkcolor=blue,citecolor=blue,urlcolor=blue}
\hypersetup
{
  pdfauthor={J. Enrique V\'azquez\,-Lozano (juavazlo@ntc.upv.es), 
  Alejandro Mart\'inez (amartinez@ntc.upv.es) \&
  Francisco J. Rodr\'iguez-Fortu\~no (francisco.rodriguez\_fortuno@kcl.ac.uk)},
  pdfsubject={Nanophotonics Technology Center [Universitat Polit\`ecnica de Val\`encia] \& 
  Department of Physics [King's College London]},
  pdftitle={Near-Field Directionality Beyond the Dipole Approximation: Electric Quadrupole and Higher-Order Multipole Angular Spectra},
  pdfkeywords={near-field directionality, spin-orbit interaction of light, quantum spin Hall effect, angular spectrum representation, Weyl's identity, multipole fields, fundamentals of nano-optics, photonics and electromagnetic theory}
}

\begin{document}
\title{Near-Field Directionality Beyond the Dipole Approximation: Electric Quadrupole and Higher-Order Multipole Angular Spectra}
\author{J. Enrique V\'azquez\,-Lozano}
\email{juavazlo@ntc.upv.es}
\affiliation{Nanophotonics Technology Center, Universitat Polit\`ecnica de Val\`encia, Camino de Vera s/n, 46022 Valencia, Spain}
\affiliation{Department of Physics, King's College London, Strand, London WC2R 2LS, United Kingdom}
\author{Alejandro Mart\'inez}
\affiliation{Nanophotonics Technology Center, Universitat Polit\`ecnica de Val\`encia, Camino de Vera s/n, 46022 Valencia, Spain}
\author{Francisco J. Rodr\'iguez-Fortu\~no}
\email{francisco.rodriguez\_fortuno@kcl.ac.uk}
\affiliation{Department of Physics, King's College London, Strand, London WC2R 2LS, United Kingdom}
\date{\today}

\begin{abstract}
Within the context of spin-related optical phenomena, the near-field directionality is generally understood from the quantum spin Hall effect of light, according to which the transverse spin of surface or guided modes is locked to the propagation direction. So far, most previous works have been focused on the spin properties of circularly polarized dipolar sources. However, in near-field optics, higher-order multipole sources (e.g., quadrupole, octupole, and so on) might become relevant, so a more in-depth formulation would be highly valuable. Building on the angular spectrum representation, we provide a general, analytical, and ready-to-use treatment in order to address the near-field directionality of any multipole field, particularizing to the electric quadrupole case. Besides underpinning and upgrading the current framework on spin-dependent directionality, our results may open up new perspectives for engineering light-matter coupling at the nanoscale.
\end{abstract}
\maketitle
\sloppy

\textit{Introduction.\textemdash}The current trend toward the miniaturization and integration of photonic devices has spurred the unprecedented ability for exploiting the spin of light in a multitude of nanophotonic applications based on the so-called spin-orbit interaction (SOI) \cite{Bliokh2015a,Cardano2015}. Essentially, optical SOI comprises a broad class of effects involving the mutual influence between the state of polarization (spin) and the spatial propagation features (orbit) of evanescent as well as nontrivially structured optical fields \cite{Liberman1992}, that naturally and necessarily emerges at the subwavelength scale \cite{VazquezLozano2018}. In classical and quantum optics, one of the most important evidences of the occurrence of SOI comes from the spin-controlled unidirectional excitation of guided waves \cite{Fortuno2013}, which has been successfully demonstrated both theoretically and experimentally in a wide variety of photonic platforms and spectral ranges, including dielectric-based integrated optics \cite{Fortuno2014a,Fortuno2014b,Fortuno2014c}, plasmonic systems \cite{OConnor2014}, hyperbolic metamaterials \cite{Kapitanova2014}, photonic crystals \cite{Feber2015,Sollner2015}, microwave waveguides \cite{Carbonell2014}, and optical fibers \cite{Petersen2014}. Within this context, the relatively recent discovery of the photonic counterpart of the quantum spin Hall effect \cite{Bliokh2015b} may be regarded as a major breakthrough on the unified understanding of the spin-momentum locking, and its incontrovertible relationship with the transverse spin associated to the evanescent waves supported by surface or guided modes \cite{Bliokh2012,Bliokh2014,Mechelen2016}. This general framework has provided an insightful explanation for the near-field directionality of electromagnetic (EM) guided modes by circularly polarized electric (and/or magnetic) dipoles \cite{Fortuno2013}. Indeed, from the point of view of the local dynamical properties, the spin-controlled unidirectional excitation can be simply understood as the coupling between the longitudinal spin components of the dipolar source with that of the corresponding guided mode \cite{Marrucci2015,Aiello2015,Espinosa-Soria2016}. 

For a more accurate description of near-field coupling a full-vector wave analysis becomes necessary \cite{Novotny}, thereby taking into account both the relative amplitude and phase between the electric and magnetic field contributions \cite{Picardi2017,Picardi2018a}. This leads to an approach that strongly resembles Fermi's golden rule, according to which the chiral (or directional) waveguide coupling efficiency is proportional to $\left|{\bf p}\cdot{\bf E}^*+{\bf m}\cdot\mu{\bf H}^*\right|^2$, i.e., it relies on the similarity between the electric (and/or magnetic) dipole moment, ${\bf p}$ (and/or ${\bf m}$), and the electric (and/or magnetic) field distribution, ${\bf E}$ (and/or ${\bf H}$) of the guided mode at the same location of the dipolar sources, with $\mu$ being the permeability of the medium \cite{Petersen2014,Feber2015,Marrucci2015,Aiello2015,Espinosa-Soria2016}. Nonetheless, in structures exhibiting a translational symmetry along two directions of a given plane, this scheme for the mode-coupling can be further simplified \cite{Picardi2017}. Indeed, in these scenarios, one can also employ an alternative and equivalent approach based on the asymmetric features of the near-field \textit{angular spectrum representation} together with considerations of structural symmetry and momentum conservation \cite{Fortuno2013,OConnor2014,Neugebauer2014,Neugebauer2016}. This formalism has the advantage of only accounting for the matching condition between the wavevector of the electric (and/or magnetic) dipole and that of the corresponding confined mode, thus gaining some physical intuition and simplifying the mathematics.

Regardless of the specific approach, up to our knowledge, all previous works addressing the unidirectional near-field scattering have only been focused on dipolar sources. This includes the Janus dipole \cite{Picardi2018a,Picardi2019}, being side-dependent topologically protected, as well as the directional Huygens' dipole \cite{Wei2018}. Yet, higher-order multipole moments (e.g., quadrupole, octupole, and so on) turn out to be relevant at the nanoscale \cite{Novotny,Keller}, and a more in-depth treatment for the spin-dependent directionality beyond the dipole approximation is required. Building upon the angular spectrum representation, in this work we provide a complete and systematic formulation for the near-field directionality of EM multipole fields of arbitrary order. Special emphasis is placed on the particular case of the electric quadrupole, for which we explicitly show the potential benefits of increasing the degrees of freedom available as well as of broadening the spatial range where the evanescent modes have a significant contribution \cite{VazquezLozano2018,Corbaton2016}. Importantly, the end results are elegantly expressed in an easy-to-use form. Indeed, we also include an online tool which directly provides the angular spectra for arbitrary multipole sources \cite{link_applet}. In this way, they can be directly applied to the analytical design of nanoscale optical sources for engineering the directional scattering and coupling of EM fields to both confining structures and waveguiding systems \cite{Curto2013,Evlyukhin2013,Alaee2018a,Alaee2018b}.

\textit{Angular spectrum of electric quadrupole.\textemdash}The angular spectrum representation, often referred to as the \textit{generalized plane wave expansion} \cite{Devaney1974}, is a very well known theoretical concept that allows us to obtain a mode representation of any EM field in terms of elementary plane waves which can be either propagating or evanescent \cite{Mandel,Novotny}. It has been found to be especially well suited for describing near- and far-field light-matter interaction of optical sources neighboring material structures exhibiting planar geometries, such as slabs, interfaces or layered media \cite{Keller,Mandel,Novotny}. Accordingly, this approach has been extensively used in a plethora of fundamental problems in classical optics (see, e.g., Refs. \cite{Friberg1983,Sherman1976,Wolf1985} and references therein), with special emphasis on the study of reflection and transmission of EM multipole fields by interfaces \cite{Inoue1998,Arnoldus2005}, and more recently for the characterization of near-field directionality of dipolar sources \cite{Fortuno2013,Picardi2017,Neugebauer2014,Neugebauer2016}. 

Regarding our main goal toward a complete description of the near-field directionality, the aforementioned angular spectrum formalism is actually used to represent the EM radiation emanating from a localized optical source via the partial Fourier transform \cite{Goodman}:
\begin{equation}
\tilde{{\bf \Psi}}(\kappa_x,\kappa_y;z)\equiv\frac{k^2}{4\pi^2}\iint_{-\infty}^{+\infty}{{\bf \Psi}({\bf r})e^{-ik\left(\kappa_x x+\kappa_y y\right)}dxdy},
\label{Eq1}
\end{equation}
with ${\bf \Psi}({\bf r})$ being the complex amplitude of any scalar or vector field satisfying the \textit{Helmholtz wave equation} so that $\tilde{{\bf \Psi}}(\kappa_x,\kappa_y;z)=\tilde{{\bf \Psi}}_0(\kappa_x,\kappa_y;0)e^{\pm ik\kappa_z z}$. For the sake of completeness as well as for comparison purposes, we will firstly sketch the derivation of the angular spectrum (momentum representation) of an electric dipole \cite{Setala1999,Mandel}. To this end, it should be noted that, in general, any optical source (i.e., dipole, quadrupole, and so on) can be characterized by either the charge-current density distribution, or alternatively through the associated vector potential \cite{Novotny}. Then, following Jackson's textbook on classical electrodynamics \cite{Jackson}, for the particular case of the electric dipole (ED), the vector potential reads as
\begin{equation}
{\bf A}^{\rm ED}=\frac{\mu}{4\pi}\left[\int_{-\infty}^{+\infty}{{\bf J}({\bf r}')d{\bf r}'}\right] \frac{e^{ikr}}{r}=-\frac{i\mu\omega}{4\pi}{\bf p}\frac{e^{ikr}}{r},
\label{Eq2}
\end{equation}
where ${\bf p}\equiv\int{{\bf r}'\rho({\bf r}')d{\bf r}'}$ is the electric dipole moment, which is in turn tied to the electric current density ${\bf J}({\bf r})=-i\omega \delta^3({\bf r}-{\bf r}'){\bf p}$, and $\rho({\bf r})$ is the charge density. Notice that throughout this work we will assume fields with harmonic time dependence of the form $e^{-i\omega t}$, where $\omega$ is the angular frequency. Making use of \textit{Weyl's identity} \cite{Weyl1919,Mandel},
\begin{equation}
\frac{e^{ikr}}{r}=\frac{ik}{2\pi}\iint_{-\infty}^{+\infty}{\frac{1}{\kappa_z}e^{ik\left(\kappa_xx+\kappa_yy\pm\kappa_zz\right)}d\kappa_xd\kappa_y},
\label{Eq3}
\end{equation}
it is straightforward to show that
\begin{equation}
{\bf A}^{\rm ED}=\frac{\mu c k^2}{8\pi^2 n}{\bf p}\iint_{-\infty}^{+\infty}{\frac{1}{\kappa_z}e^{ik\left(\kappa_xx+\kappa_yy\pm \kappa_zz\right)}d\kappa_xd\kappa_y},
\label{Eq4}
\end{equation}
where ${\bf k}^{\pm}=k\left(\kappa_x,\kappa_y,\pm\kappa_z\right)$ is the wavevector, and the signs $+$ and $-$ stand for the wave propagation through the half-spaces $z>0$ and $z<0$, respectively. Furthermore, solutions to the Helmholtz wave equation must obey ${\bf k}^{\pm}\cdot{\bf k}^{\pm}=k^2$, so it follows that
\begin{equation}
\kappa_z=\left\{\begin{matrix}
\displaystyle\sqrt{1-\kappa_x^2-\kappa_y^2},&\;\text{if}\;& \kappa_x^2+\kappa_y^2\leq 1, \\
\displaystyle i\sqrt{\kappa_x^2+\kappa_y^2-1},&\;\text{if}\;& \kappa_x^2+\kappa_y^2>1,
\end{matrix}\right.
\label{Eq5}
\end{equation}
which correspond, respectively, to the propagating and evanescent modes in the partial Fourier (or momentum) space $(\kappa_x,\kappa_y)$ \cite{Mandel,Novotny}. Hence, according to the basic properties of Fourier transform, from Eqs. \eqref{Eq1} and \eqref{Eq4} the spectral amplitude of the electric dipole vector potential is $\tilde{{\bf A}}^{\rm ED}(\kappa_x,\kappa_y;z)=\tilde{{\bf A}}_0^{\rm ED}(\kappa_x,\kappa_y;0)e^{\pm ik\kappa_zz}$, where
\begin{equation}
\tilde{{\bf A}}_0^{\rm ED}(\kappa_x,\kappa_y;0)=\frac{\mu c k^2}{8\pi^2 n}\frac{1}{\kappa_z}{\bf p}.
\label{Eq6}
\end{equation}

Following a similar but slightly trickier procedure we may obtain a closed expression for the angular spectrum amplitude of the electric quadrupole as well. To this aim, we start again from the vector potential of the electric quadrupole (EQ) as given in Ref. \cite{Jackson}:
\begin{equation}
{\bf A}^{\rm EQ}=\frac{\mu\omega k}{24\pi}\left(\frac{1}{ikr}-1\right)\left[\dvec{\mathcal{Q}}\cdot{\bf e}_r\right]\frac{e^ {ikr}}{r},
\label{Eq7}
\end{equation}
where $\dvec{\mathcal{Q}}\cdot{\bf e}_r$ is the term that conveys the ``quadrupolar character'' to the vector potential, with ${\bf e}_r$ being the radial unit vector. The quadrupole moment $\dvec{\mathcal{Q}}$ is a rank-two tensor, usually represented by a $3\times3$ traceless, complex, and symmetric matrix, thus reducing the unknowns to $5$. In this case the scalar contribution of Eq. \eqref{Eq7} is no longer a proper solution of the Helmholtz wave equation. Moreover, since ${\bf e}_r$ depends on the direction of observation, we cannot take it out of the integral \cite{SupplementalMaterial}. With these considerations in mind, the spectral amplitude of the electric quadrupole vector potential can be directly obtained by applying a partial Fourier transform as defined in Eq. \eqref{Eq1} to the vector potential Eq. \eqref{Eq7}. In this way, after some straightforward but lengthy manipulations that involves the change of variables to cylindrical coordinates (in both real and momentum space), the utilization of some integral identities leading to Bessel functions \cite{Watson}, and a judicious choice of integration by parts (see Sec. III in the Supplemental Material \cite{SupplementalMaterial}), we finally find that, in the limit $z\to 0$,
\begin{equation}
\tilde{{\bf A}}_0^{\rm EQ}(\kappa_x,\kappa_y;0)=\frac{\mu c k^2}{8\pi^2n}\frac{1}{\kappa_z}\frac{(-i)\left[\dvec{\mathcal{Q}}\cdot{\bf k}^{\pm}\right]}{6}.
\label{Eq8}
\end{equation}
When performing the integration, it is important to take into account the divergent behavior at the origin due to the presence of a singularity. This also happens for the electric dipole \cite{Mandel}, and may lead to misleading outcomes. Despite that, as pointed out in the Supplemental Material \cite{SupplementalMaterial}, we can formally obtain the angular spectrum of the electric quadrupole in each of the half-spaces [Eq. \eqref{Eq8}].

Equation \eqref{Eq8}, expressed analytically in a closed form, constitutes the first main result of this work. In this respect, it is worth pointing out that it has been deliberately written in such a way that is easy to compare it with the angular spectrum amplitude of the electric dipole. Indeed, one can directly get the spectral amplitude of the electric quadrupole [Eq. \eqref{Eq8}] from that of the electric dipole [Eq. \eqref{Eq6}] simply by substituting ${\bf p}\to (-i)\left[\dvec{\mathcal{Q}}\cdot{\bf k}^{\pm}\right]/6$. Furthermore, these results are perfectly consistent with those presented in Ref. \cite{Evlyukhin2013} for the multipole decomposition of scattered fields by arbitrary-shaped nanoparticles in the far-field approximation. According to this, the light-induced polarization vector can be Taylor expanded as follows
\begin{equation}
{\bf P}({\bf r})\approx\left[{\bf p}-\frac{1}{6}\dvec{\mathcal{Q}}\nabla+\frac{i}{\omega}\left(\nabla\times{\bf m}\right)+\ldots\right]\delta({\bf r}).
\label{Eq9}
\end{equation}
It should be noted that ${\bf P}$ is related to the electric current density ${\bf J}$, which is in turn tied to the vector potential ${\bf A}$ \cite{Novotny,Jackson}. Thus, by means of the following substitution $\nabla\to i{\bf k}^{\pm}$, it can be readily seen that Eq. \eqref{Eq9} coincides with Eqs. \eqref{Eq6} and \eqref{Eq8}, as well as with the magnetic dipole term (see appendix in the Supplemental Material \cite{SupplementalMaterial}):
\begin{eqnarray}
\!\!\!\!\!\!\nonumber \tilde{{\bf A}}_0&=&\tilde{{\bf A}}_0^{\rm ED}+\tilde{{\bf A}}_0^{\rm EQ}+\tilde{{\bf A}}_0^{\rm MD}\\
&=&\frac{\mu ck^2}{8\pi^2n}\frac{1}{\kappa_z}\left[{\bf p}-\frac{i}{6}\left(\dvec{\mathcal{Q}}\cdot{\bf k}^{\pm}\right)-\frac{1}{\omega}\left({\bf k}^{\pm}\times{\bf m}\right)\right].
\label{Eq10}
\end{eqnarray}
Besides providing a rapid verification of the above results, this correspondence between $\tilde{{\bf A}}_0$ and ${\bf P}$ permits us to infer the next higher-order terms through a simple identification of each term in the Taylor expansion of the Dirac delta function around the origin \cite{footnote_far-field_approx}.

Practical applications usually require the angular spectrum of the EM fields themselves instead of that of the vector potential \cite{Fortuno2013,Picardi2017,Picardi2018a}. In fact, this is a major advantage of the angular spectrum approach, because cumbersome analysis involving the computation of gradients, divergences, curls, and Laplacian, is reduced to simple algebra relying upon the product between the wavevector ${\bf k}^{\pm}$ and the spectral amplitudes $\tilde{{\bf A}}_0$. Specifically, from the relationship between the magnetic field and the vector potential, ${\bf B}=\mu{\bf H}=\nabla\times{\bf A}$, together with the curl-like Maxwell equation $\nabla\times{\bf H}=-i\omega\varepsilon{\bf E}$, with $\varepsilon$ being the permittivity of the medium, the complex field amplitudes in momentum space $(\kappa_x,\kappa_y)$ can be nicely expressed in a compact form:
\begin{eqnarray}
&&\!\!\!\!\!\!\!\!\!\!\tilde{\bf E}(\kappa_x,\kappa_y;0)\!=\!\frac{ik^3}{8\pi^2\varepsilon}\frac{\left\{[{\bf e}_s\cdot {\bf p}_{\rm eff}]{\bf e}_s+[{\bf e}_p^\pm\cdot{\bf p}_{\rm eff}]{\bf e}_p^\pm\right\}}{\kappa_z},
\label{Eq11}\\
&&\!\!\!\!\!\!\!\!\!\!\tilde{\bf H}(\kappa_x,\kappa_y;0)\!=\!\frac{ik^3}{8\pi^2\varepsilon }\frac{1}{Z}\frac{\left\{[{\bf e}_p^\pm\cdot {\bf p}_{\rm eff}]{\bf e}_s-[{\bf e}_s\cdot{\bf p}_{\rm eff}]{\bf e}_p^{\pm}\right\}}{\kappa_z},
\label{Eq12}
\end{eqnarray}
with ${\bf p}_{\rm eff}$ corresponding to the term in square brackets in Eq. \eqref{Eq10}, and $Z=\sqrt{\mu/\varepsilon}$ is the medium impedance. These expressions are readily obtained by introducing the polarization vector basis, ${\bf e}_s=\left(-\kappa_y,\kappa_x,0\right)/\kappa_R$ and ${\bf e}_p=\left(\pm\kappa_x\kappa_z,\pm \kappa_y\kappa_z,-\kappa_R^2\right)/\kappa_R$, where the subscripts indicate the $s$ (or TE) and $p$ (or TM) polarizations, and $\kappa_R\equiv(\kappa_x^2+\kappa_y^2)^{1/2}$ \cite{SupplementalMaterial}. Notice that in the lossless propagating case (i.e., when ${\bf k}^{\pm}$ is real), they correspond to the usual unit vectors in spherical coordinates ${\bf e}_\varphi$ and ${\bf e}_\theta$, respectively. 

\begin{figure}[t!]
\includegraphics[width=0.9\linewidth]{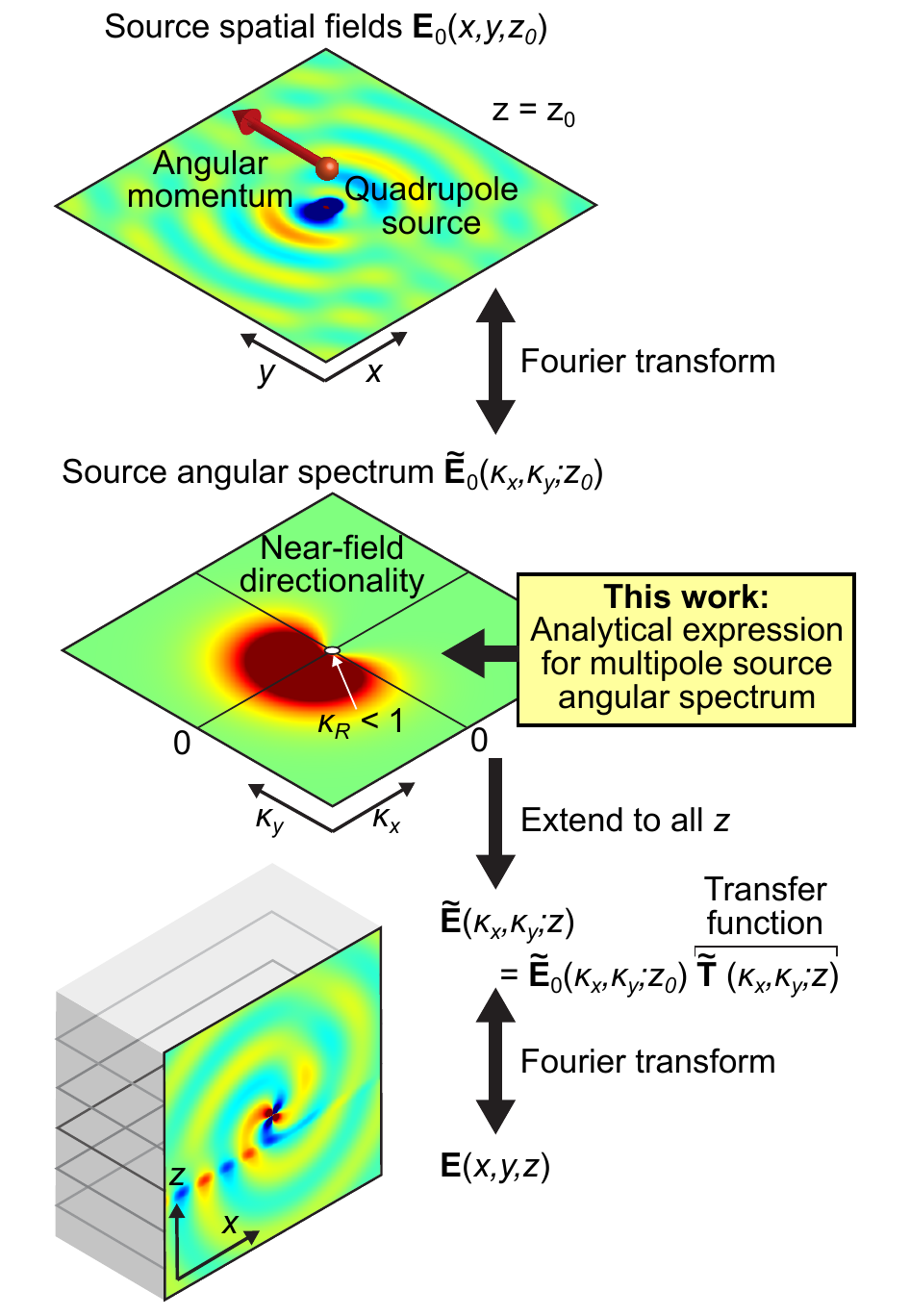}
\caption{Schematic description for visualizing the near-field directionality of arbitrary multipole sources. Starting from the space-dependent complex representation for the EM fields, over a given plane, e.g., $z=z_0$, the angular spectrum is directly obtained via the partial Fourier transform [Eq. \eqref{Eq1}]. Within this representation, the analytic extension to the whole space with a dielectric slab is simply performed by using a transfer function, depending on the Fresnel's coefficients.}
\label{fig_01}
\end{figure}

\textit{Near-field directionality beyond the dipole.\textemdash}In translationally symmetric nanostructures, such as dielectric slab waveguides or metal-dielectric interfaces supporting surface plasmons, the near-field directional coupling of guided modes can be simply understood either from the matching condition of the local EM fields (underpinning the aforementioned Fermi's golden rule) \cite{Feber2015,Petersen2014,Marrucci2015,Aiello2015,Espinosa-Soria2016} or, alternatively, from the angular spectrum representation accounting for the phase-matching condition \cite{Picardi2017,Picardi2018a}. Whereas the former approach requires prior knowledge of the EM field structure of the guided modes, the angular spectrum formalism only relies on the source by itself, together with structural symmetry aspects, thereby linked to momentum conservation \cite{Fortuno2013,OConnor2014,Neugebauer2014,Neugebauer2016}. Hence, it can also be applied to nonplanar structures as long as they are translationally invariant along a given direction \cite{Picardi2017}.

The results given in Eqs. \eqref{Eq11} and \eqref{Eq12} correspond to the angular spectrum of the source, which are equal to the angular spectrum of the total field in an unbounded homogeneous medium. In the presence of nearby structures with translational invariance in the $x$ and $y$ directions, the reflected and transmitted fields can be easily included. Indeed, their angular spectra is simply given by the angular spectrum of the source with a multiplicative transfer function \cite{Mandel}, as shown in Fig. \hyperref[fig_01]{1}. In the case of a single interface, the transfer function simply involves reflection and transmission Fresnel's coefficients \cite{Novotny}. Therefore, the near-field directional coupling of the source depends solely on the asymmetry in the evanescent components of the angular spectrum of the optical sources. This fact has been shown for circularly polarized electric and magnetic dipoles \cite{Picardi2017}, as well as for Huygens' and Janus sources \cite{Picardi2018a,Picardi2019}. The same idea can of course be exploited for the electric quadrupole case [see Fig. \hyperref[fig_02]{2(b)}]. Indeed, as follows from Eqs. \eqref{Eq10}-\eqref{Eq12}, the electric quadrupole with non-zero elements $\mathcal{Q}_{11}=-\mathcal{Q}_{33}=-1$ and $\mathcal{Q}_{13}=-\mathcal{Q}_{31}=i$ shows a strongly asymmetrical $p$-polarized angular spectrum, with high amplitude in the near-field region $\kappa_x<-1$, and negligible amplitude for $\kappa_x>1$, as shown in Fig. \hyperref[fig_01]{1}, thus behaving very similarly to a circularly polarized electric dipole source. As a consequence, both sources show a clear unidirectionality when placed near a waveguide or slab [see Figs. \hyperref[fig_02]{2(a-b)}]. Importantly, a quadrupole source has more degrees of freedom than a dipole, due to its tensorial nature, thereby allowing a more versatile engineering of its near-field for potential applications.

\begin{figure*}[t!]
\includegraphics[width=15.5cm,keepaspectratio]{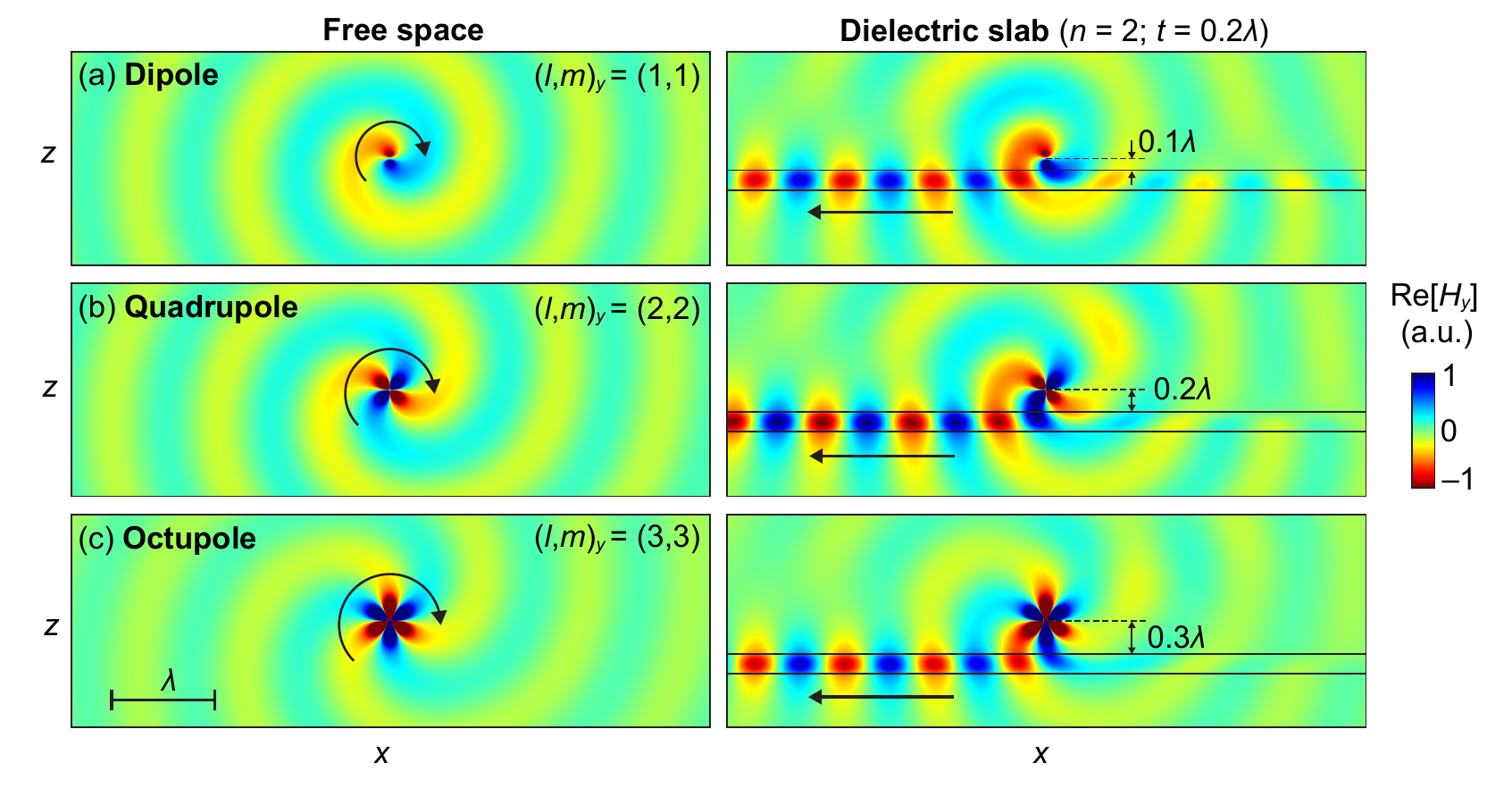}
\caption{Electromagnetic field calculation of electric multipolar sources in free space (left) and near a dielectric slab waveguide (right), showing directional excitation of TM guided modes. The analytical angular spectra of multipolar sources was combined with a transfer matrix method to obtain the angular spectra at various $z$ planes, and fields are calculated using numerical inverse fast Fourier transform on each plane. (a) Circularly polarized electric dipole ${\bf p} = (1,0,i)$ corresponding to electric multipole $(l,m)_y=(1,1)$ where the subscript $y$ denotes the angular momentum axis oriented towards $y$. (b) ``Circularly'' polarized electric quadrupole with non-zero elements $\mathcal{Q}_{11}=-1, \mathcal{Q}_{13}=i, \mathcal{Q}_{31}=i$ and $\mathcal{Q}_{33}=1$, corresponding to electric multipole $(l,m)_y=(2,2)$. (c) ``Circularly'' polarized electric octupole corresponding to electric multipole $(l,m)_y=(3,3)$.}
\label{fig_02}
\end{figure*}

\textit{Angular spectrum of higher-order multipole fields.\textemdash}In an experimentally realistic situation, optical sources are actually emitters (quantum dots or atoms) or scatterers (single or assemblies of nanoparticles). In both cases, the multipole expansion of EM fields, expressed in terms of the \textit{vector spherical harmonics} \cite{Jackson}, afford a suitable tool, because they enable a complete description of fields emanating from or coupling to localized optical systems \cite{Novotny}. This is crucial in nanophotonics, typically dealing with strongly confined optical fields that may lead to steep field gradients. Hence, it should be highly valuable for the near-field directional scattering when accounting for higher-order spectroscopic resonances \cite{Alaee2018a,Alaee2018b}.

In this case, instead of considering the standard vector potentials, in order to treat the electric and magnetic multipole fields on an equal footing, we will use a formulation based on the Hertz potentials \cite{Essex1977,Ornigotti2014}. Under this representation, the magnetic and electric fields are, respectively, related to the electric- and magnetic-like Hertz potentials \cite{footnote_source}, in the following manner:
\begin{equation}
{\bf H}_{l,m}^{\rm (e)}=-i\omega[\nabla\times{\bf \Pi}_{l,m}^{\rm (e)}],\quad\; {\bf E}_{l,m}^{\rm (m)}=i\mu_0\omega[\nabla\times{\bf \Pi}_{l,m}^{\rm (m)}].
\label{Eq13}
\end{equation}
After some manipulations taking into account the definitions of the EM multipole fields and the fulfillment of Helmholtz wave equation \cite{Jackson}, one can find a closed expression for the Hertz vector potential of any electric and magnetic multipolar source of $(l,m)$ order (see Sec. V in the Supplemental Material \cite{SupplementalMaterial} for details). Finally, from the partial Fourier transform, it can be proven that the spectral amplitude of the Hertz potentials associated to the electric-like sources on the plane $z=0$ reads as
\begin{eqnarray}
\tilde{\bf \Pi}^{\rm (e)}_{l,m}=\frac{-k}{4\pi\omega}\frac{(-i)^mC_{l,m}}{\sqrt{l(l+1)}}\left[\tilde{\Pi}_{l,m}^{\rm (e),x},\tilde{\Pi}_{l,m}^{\rm (e),y},\tilde{\Pi}_{l,m}^{\rm (e),z}\right],
\label{Eq14}
\end{eqnarray}
with
\begin{eqnarray*}
&&\tilde{\Pi}_{l,m}^{\rm(e),x}=i\left[(m-l)\tilde{\mathcal{K}}_{l,m+1}^{l,m}e^{i\phi}+(m+l)\tilde{\mathcal{K}}_{l,m-1}^{l,m}e^{-i\phi}\right]e^{im\phi},\\
&&\tilde{\Pi}_{l,m}^{\rm(e),y}=\left[(m-l)\tilde{\mathcal{K}}_{l,m+1}^{l,m}e^{i\phi}-(m+l)\tilde{\mathcal{K}}_{l,m-1}^{l,m}e^{-i\phi}\right]e^{im\phi},\\
&&\tilde{\Pi}_{l,m}^{\rm(e),z}=2(l+m)\tilde{\mathcal{K}}_{l,m}^{l-1,m}e^{im\phi}.
\end{eqnarray*}
Same results also hold for magnetic-like sources, just by noting that $\tilde{\bf \Pi}^{\rm (m)}_{l,m}=-c\sqrt{\mu_r/\varepsilon_r}\tilde{\bf \Pi}^{\rm (e)}_{l,m}$ \cite{Ishimaru}. These expressions depend on the function $\mathcal{K}_{l,m}^{l',m'}$ in the limit $z\to 0$ \cite{footnote_limit}:
\begin{equation}
\tilde{\mathcal{K}}_{l,m}^{l',m'}(k,\kappa_R)\equiv\lim\limits_{z\to0}{P_{l'}^{m'}(z)\mathcal{I}_{l,m}(k,\kappa_R;z)},
\label{Eq15}
\end{equation}
where $P_{l'}^{m'}$ are the associated Legendre polynomials of degree $l'$ and order $m'$, and
\begin{eqnarray}
&&\nonumber\mathcal{I}_{l,m}(k,\kappa_R;0)=\frac{\sqrt{\pi}\kappa_R^m}{k^2}\left[\frac{\Gamma\left[\frac{1+l+m}{2}\right]}{\Gamma\left[\frac{l-m}{2}\right]}+i\left(-1\right)^l\frac{\Gamma\left[\frac{2-l+m}{2}\right]}{\Gamma\left[\frac{1-l-m}{2}\right]}\right]\\
&&\quad \quad\frac{{}_2F_1\left[\frac{1}{2}\left(2-l+m\right),\frac{1}{2}\left(1+l+m\right),1+m,\kappa_R^2\right]}{\Gamma\left[1+m\right]},
\label{Eq16}
\end{eqnarray}
with ${}_2F_1(a,b,c;z)$ being the Gaussian hypergeometric function \cite{Abramowitz}. Equation \eqref{Eq16} is the second main result of this work; it can be regarded as a generalization of Weyl's identity, in the sense that it enables the calculation of the angular spectrum representation of any EM multipole field. It should be noted that a similar result was previously obtained \cite{Bobbert1986,Borghi2004}, but using a very different notation \cite{footnote_applet}. Importantly, as shown in Sec. VI of the Supplemental Material \cite{SupplementalMaterial}, apart from the prefactors, these analytical and closed forms are in perfect agreement with the particular cases of the electric dipole ($l=1$) [Eq. \eqref{Eq6}] and quadrupole ($l=2$) [Eq. \eqref{Eq8}], thereby confirming their validity, at least up to the electric quadrupole case. 

The circularly polarized dipole and quadrupole shown in Figs \hyperref[fig_02]{2(a-b)} may be expressed in multipole form as $(l,m)_y = (1,1)$ and $(2,2)$ respectively. The $y$ subindex corresponds to the angular momentum axis of the source, therefore they correspond to dipolar and quadrupolar sources with the highest angular momentum along $y$, i.e., parallel to the interface. This sets up a circular phase gradient which sweeps the surface of the waveguide along $-x$, causing unidirectional excitation. In Fig \hyperref[fig_02]{2(c)} we confirm the trend by considering the $(l,m)_y = (3,3)$ octupole source. A clear advantage of higher order sources is their longer range for near-field directional coupling, as multipoles of increasing order result in a similar amplitude of guided mode excitation at increasing distances from the waveguiding structure (for the same far-field radiated amplitude) as shown in Fig \hyperref[fig_02]{2}. This is reflected in the analytical angular spectra as multiplicative ``$\kappa_R$'' terms, which enhance the amplitude of the higher frequencies, thus broadening the spectrum. Notice that our analytical Eqs. \eqref{Eq14}-\eqref{Eq16} assume $(l,m)_z$ sources, whose angular momentum is around the $z$ axis and do not show near-field directionality in the $XY$ plane, but the same expression can be used to compute the angular spectra of the $(l,m)_y$ sources used in Fig \hyperref[fig_02]{2} by properly rotating the $\kappa_x$, $\kappa_y$ and $\kappa_z$ axes. Our online spectrum calculator allows this selection \cite{link_applet}.

\textit{Summary and outlook.\textemdash}Taking advantage of the angular spectrum representation we have presented a general, systematic, and ready-to-use formulation for the near-field directionality beyond the dipole approximation. Additionally, for the readers' convenience, we include an online tool to retrieve the angular spectra of arbitrary multipoles \cite{link_applet}. Specifically, we have derived an analytical and closed expression for the angular spectrum of EM multipole fields of arbitrary order. We have seen that successively higher order multipole fields provide an increasing range of near-field directionality as well as an increasing number of degrees of freedom. This could be useful, for instance, to promote the spin-controlled (coherent) multidirectional excitation of guided waves. Hence, besides underpinning and upgrading the already known framework on near-field directionality \cite{Fortuno2013,Fortuno2014a,Fortuno2014b,Fortuno2014c,Fortuno2015,Espinosa-Soria2017,Feber2015,Petersen2014,OConnor2014,Carbonell2014,Kapitanova2014,Sollner2015}, the full consideration of higher-order multipole moments may unveil new spin-dependent features, specially in the context of nonlinear nanophotonics \cite{Smirnova2016}.

This work was supported by funding from Ministerio de Econom\'ia y Competitividad (MINECO) of Spain under Contract No. TEC2014-51902-C2-1-R and by European Research Council Starting Grant ERC-2016-STG-714151-PSINFONI. Also, J.E.V.-L. thanks the kind hospitality of Department of Physics, King's College London, where this research was partially carried out.

\newpage
\onecolumngrid
\newpage
\begin{center}
{\large \textbf{Supplemental Material:\\Near-Field Directionality Beyond the Dipole Approximation: Electric Quadrupole and Higher-Order Multipole Angular Spectra}}
\end{center}
\begin{center}
J. Enrique V\'azquez\,-Lozano,$^{1,2,\blue{*}}$ Alejandro Mart\'inez,$^{1}$ and Francisco J. Rodr\'iguez-Fortu\~no$^{2,\blue{\dagger}}$\\
$^{1}${\small\textit{Nanophotonics Technology Center, Universitat Polit\`ecnica de Val\`encia, Camino de Vera s/n, 46022 Valencia, Spain}}\\
$^{2}${\small\textit{Department of Physics, King's College London, Strand, London WC2R 2LS, United Kingdom}}\\
(Dated: \today)
\end{center}
\begin{quotation}
In this supplemental material we provide a step-by-step derivation of the angular spectrum representation of any electromagnetic multipole field. For completeness, and for convenience throughout this work, we first briefly review the angular spectrum of the electric dipole, which relies on Weyl's identity. In a similar but slightly trickier way, we are also able to get the angular spectrum of the electric quadrupole. Due to the similarity between these two particular results, we look into the possibility of addressing a general approach valid to any multipolar order. In this way, we find an analytical and closed expression that generalizes Weyl's identity thus enabling the angular spectrum representation of any multipole field.
\end{quotation}
\renewcommand\thepage{S.\arabic{page}}
\setcounter{page}{1}
\renewcommand\theequation{S\arabic{equation}}
\setcounter{equation}{0}
\maketitle
\sloppy
\section{I. Angular spectrum representation, Weyl's identity and Hertz potentials}

The angular spectrum representation is a classical technique that allows us to express any electromagnetic (EM) field in homogeneous media as a superposition of elementary plane waves which can be propagating (homogeneous) or evanescent (inhomogeneous) \cite{Mandelsm,Novotnysm}. Within the dipole approximation \cite{Setala1999sm} this formalism essentially relies on \textit{Weyl's identity} \cite{Weyl1919sm}:
\begin{equation}
\frac{e^{ikr}}{r}=\frac{ik}{2\pi}\iint_{-\infty}^{+\infty}{\frac{1}{\kappa_z}e^{ik\left(\kappa_xx+\kappa_yy\pm\kappa_zz\right)}d\kappa_xd\kappa_y},
\label{EqS1}
\end{equation}
where the wavevector is defined as ${\bf k}^{\pm}=k\left(\kappa_x,\kappa_y,\pm\kappa_z\right)$, and the signs $+$ and $-$ stand for the wave propagation through the half-spaces $z>0$ and $z<0$, respectively. Moreover, assuming that ${\bf k}^{\pm}\cdot{\bf k}^{\pm}=k^2$, it follows that
\begin{equation}
\kappa_z=\left\{\begin{matrix}
\sqrt{1-\kappa_x^2-\kappa_y^2},&\;\text{if}\;& \kappa_x^2+\kappa_y^2\leq 1; \\
i\sqrt{\kappa_x^2+\kappa_y^2-1},&\;\text{if}\;& \kappa_x^2+\kappa_y^2>1.
\end{matrix}\right.
\label{EqS2}
\end{equation}
This fundamental result allows us to represent any diverging spherical wave as a superposition of plane waves in the partial Fourier (or momentum) space $(\kappa_x,\kappa_y)$ \cite{Goodmansm}. As it will be shown below, beyond this simple case concerning scalar waves, the importance of the Weyl's identity becomes much more evident when studying the propagation of properly defined EM vector waves. 

Let us consider a time-harmonic optical field, ${\bf \Psi}({\bf r},t)={\bf \Psi}_0({\bf r})e^{-i\omega t}$, propagating in a linear, homogeneous, isotropic, and source-free medium. The time-independent complex amplitude ${\bf \Psi}_0$ must satisfy the Helmholtz wave equation:
\begin{equation}
\left(\nabla^2+k^2\right){\bf \Psi}_0({\bf r})=0,
\label{EqS3}
\end{equation}
where ${\bf \Psi}_0$ can be either the electric or the magnetic field, $k\equiv nk_0$ is the wave number of the medium, $k_0\equiv\omega/c$ is the wave number in vacuum, and $n\equiv\sqrt{\varepsilon_r\mu_r}$ is the refractive index, being $\varepsilon_r$ and $\mu_r$ the corresponding relative permittivity and permeability of the medium. If we assume that there exists a translational symmetry along two directions of a plane, we can define a direction ${\bf r}_\perp$ which is perpendicular to this plane of symmetry, and then ${\bf \Psi}_0$ can be represented by means of the inverse spatial Fourier transform on the partial space spanned by ${\bf r}_\parallel$ as
\begin{equation}
{\bf \Psi}_0({\bf r})=\iint_{-\infty}^{+\infty}{\tilde{{\bf \Psi}}_0({\bf k}_\parallel;{\bf r}_\perp)e^{i{\bf k_\parallel}\cdot{\bf r}_\parallel}d{\bf k}_\parallel},
\label{EqS4}
\end{equation}
where we have expressed both position and wave vectors in terms of their projections parallel and perpendicular to the plane of translational symmetry. Substituting Eq. \eqref{EqS4} into \eqref{EqS3} it is straightforward to demonstrate that
\begin{equation}
\iint_{-\infty}^{+\infty}{\left[\nabla^2_\perp\tilde{{\bf \Psi}}_0({\bf k}_\parallel;{\bf r}_\perp)+\left(k^2-{\bf k}_\parallel\cdot{\bf k}_\parallel\right)\tilde{{\bf \Psi}}_0({\bf k}_\parallel;{\bf r}_\perp)\right]e^{i{\bf k}_\parallel\cdot{\bf r}_\parallel}d{\bf k}_\parallel}=0.
\label{EqS5}
\end{equation}
For simplicity, and without loss of generality, we particularize to the case in which ${\bf k}_\parallel=k\left(\kappa_x,\kappa_y,0\right)$, and ${\bf k}_\perp^\pm=k\left(0,0,\pm\kappa_z\right)$, and so, it is easy to realize that the expression in the square brackets of Eq. \eqref{EqS5} actually is a Helmholtz-like equation reduced to the one-dimensional case:
\begin{equation}
\frac{\partial^2\tilde{{\bf \Psi}}_0(\kappa_x,\kappa_y;z)}{\partial z^2}+k^2\kappa_z^2\tilde{{\bf \Psi}}_0(\kappa_x,\kappa_y;z)=0.
\label{EqS6}
\end{equation}
Hence, the angular spectrum amplitude $\tilde{{\bf \Psi}}_0$ can be simply expressed as 
\begin{equation}
\tilde{{\bf \Psi}}_0(\kappa_x,\kappa_y;z)=\tilde{{\bf \Psi}}_0(\kappa_x,\kappa_y;0)e^{\pm ik\kappa_z z}.
\label{EqS7}
\end{equation}
Finally, in order to get the \textit{angular spectrum representation} we only have to insert this result in Eq. \eqref{EqS4}:
\begin{equation}
{\bf \Psi}_0({\bf r})=\iint_{-\infty}^{+\infty}{\tilde{{\bf \Psi}}_0(\kappa_x,\kappa_y;0)e^{ ik\left(\kappa_xx+\kappa_yy\pm\kappa_z z\right)}d\kappa_xd\kappa_y},
\label{EqS8}
\end{equation}
where, once again, the signs $+$ and $-$ refer to a wave propagating in each of the half-spaces $z>0$ and $z<0$, respectively. Thus, provided that there exists a well defined translation symmetry, e.g., along the plane $(x,y)$, one can always describe the whole EM field just from looking at the field over one of these planes, let us say the plane $z=0$. As previously anticipated, this characterization is very useful because allows us to obtain a \textit{mode representation} of the optical field in terms of propagating and evanescent plane waves \cite{Mandelsm}. Specifically, these characteristic solutions can be easily distinguished in the transverse momentum space $(\kappa_x,\kappa_y)$ taking into account their relative position with respect to the unit circumference $\kappa_x^2+\kappa_y^2=1$ \cite{Novotnysm}. In this manner, from Eqs. \eqref{EqS2} and \eqref{EqS7} we see that $\kappa_x^2+\kappa_y^2<1$ leads to oscillating functions in $z$, thereby representing propagating plane waves. Otherwise, $\kappa_x^2+\kappa_y^2>1$ gives rise to exponentially decaying solutions, which are in turn identified with evanescent waves.   

Under the premises so far established, the above scheme is completely general, and can be applied to any scalar or vector field as long as it satisfies the Helmholtz wave equation. This enables a systematic analysis of the angular spectrum of any localized optical source (i.e., the electric dipole, quadrupole, and in general any higher-order multipole moment) by means of their corresponding vector potentials. More precisely, we can make use of the so-called electric and magnetic Hertz potentials \cite{Essex1977sm,Ornigotti2014sm}, which might be introduced in terms of the standard scalar and vector potentials, $\varPhi({\bf r},t)$ and $\boldsymbol{\mathcal{A}}({\bf r},t)$ respectively, as follows \cite{Jacksonsm}
\begin{equation}
\boldsymbol{\mathcal{A}}({\bf r},t)=\mu\partial_t\boldsymbol{\varPi}^{\rm (e)}({\bf r},t)+\mu_0[\nabla\times\boldsymbol{\varPi}^{\rm (m)}({\bf r},t)], \qquad\qquad \varPhi({\bf r},t)=-\frac{1}{\varepsilon}\nabla\cdot\boldsymbol{\varPi}^{\rm (e)}({\bf r},t),
\label{EqS9}
\end{equation}
where $\varepsilon\equiv\varepsilon_0\varepsilon_r$ and $\mu\equiv\mu_0\mu_r$, with $\varepsilon_0$ and $\mu_0$ being the permittivity and permeability of free space, respectively. It is interesting to point out that the superscript refers to the nature of the source that originates the EM radiation, that may be either electric or magnetic \cite{footnote1sm}. Furthermore, it should be noted that these definitions arise directly from the relationships between the time-dependent EM fields and the gauge standard potentials, $\boldsymbol{\mathcal{E}}({\bf r},t)\equiv -\nabla \varPhi({\bf r},t)-\partial_t\boldsymbol{\mathcal{A}}({\bf r},t)$ and $\boldsymbol{\mathcal{B}}({\bf r},t)\equiv\nabla\times\boldsymbol{\mathcal{A}}({\bf r},t)$, together with the Lorenz condition, $\nabla\cdot\boldsymbol{\mathcal{A}}({\bf r},t)+\varepsilon_0\mu_0\partial_t\varPhi({\bf r},t)=0$ (a complete derivation can be found in Ref. \cite{Jacksonsm}). All in all, for our purposes the key point to bear in mind is that both the electric and magnetic Hertz potentials, $\boldsymbol{\varPi}^{\rm (e)}({\bf r},t)$ and $\boldsymbol{\varPi}^{\rm (m)}({\bf r},t)$, by construction, satisfy the vector Helmholtz wave equation, and are susceptible of an expansion into their angular spectrum.

The latter relations between the time-dependent EM fields and the potentials can be applied to both electric and magnetic-like multipole sources. For the sake of clarity and simplicity, and without any loss of generality, hereinafter it will be assumed fields with harmonic time dependence of the form $e^{-i\omega t}$. Then, from the definitions given in Eq. \eqref{EqS9}, the complex field amplitudes read as
\begin{eqnarray}
{\bf E}({\bf r})&=&{\bf E}^{\rm (e)}({\bf r})+{\bf E}^{\rm (m)}({\bf r})=\frac{1}{\varepsilon}\left\{k^2{\bf \Pi}^{\rm (e)}({\bf r})+\nabla[\nabla\cdot {\bf \Pi}^{\rm (e)}({\bf r})]\right\}+i\mu_0\omega[\nabla\times{\bf \Pi}^{\rm (m)}({\bf r})],
\label{EqS10}\\
{\bf H}({\bf r})&=&{\bf H}^{\rm (e)}({\bf r})+{\bf H}^{\rm (m)}({\bf r})=-i\omega[\nabla\times{\bf \Pi}^{\rm (e)}({\bf r})]+\frac{1}{\mu_r}\left\{k^2{\bf \Pi}^{\rm (m)}({\bf r})+\nabla[\nabla\cdot{\bf \Pi}^{\rm (m)}({\bf r})]\right\}.
\label{EqS11}
\end{eqnarray}
It is worth remarking that there exist some other definitions for the Hertz potentials that may yield simpler expressions for the EM fields. In particular, in Ref. \cite{Ishimarusm} it can be found that the electric and magnetic Hertz potentials are presented separately, thus leading to EM fields associated to electric- and magnetic-like sources that are related to each other just through the \textit{duality principle} and the symmetry of Maxwell's equations. Despite that, in order to build on a sound and self-consistent framework, we will mainly focus on the approach provided in Ref. \cite{Jacksonsm}, and so we will regard the potentials as given in Eq. \eqref{EqS9}. In this respect, it is important to recall that owing to the gauge freedom, we can always find out a gauge function $\chi({\bf r},t)$ enabling the transformation of any scalar and vector potential so as to meet the Lorenz condition, and thereby giving rise to Hertz-like vector potentials. This fact will be profitably exploited later in dealing with the angular spectrum representation of the EM multipole fields.

Taking into account the appropriateness of the Hertz potentials for characterizing any multipole field, the main advantage of the angular spectrum formalism becomes evident when representing the EM fields, because the previous expressions involving spatial derivatives are greatly simplified in the partial Fourier space:
\begin{eqnarray}
\tilde{\bf E}^{\rm (e)}(\kappa_x,\kappa_y;z)&=&\frac{1}{\varepsilon}\left\{k^2\tilde{{\bf \Pi}}^{\rm (e)}(\kappa_x,\kappa_y;z)-{\bf k}^{\pm}[{\bf k}^ {\pm}\cdot \tilde{{\bf \Pi}}^{\rm (e)}(\kappa_x,\kappa_y;z)]\right\},
\label{EqS12}\\
\tilde{\bf H}^{\rm (e)}(\kappa_x,\kappa_y;z)&=&\frac{ck}{n}\left[{\bf k}^{\pm}\times\tilde{{\bf \Pi}}^{\rm (e)}(\kappa_x,\kappa_y;z)\right],
\label{EqS13}\\
\tilde{\bf E}^{\rm (m)}(\kappa_x,\kappa_y;z)&=&-\frac{Z_0k}{n}\left[{\bf k}^{\pm}\times\tilde{{\bf \Pi}}^{\rm (m)}(\kappa_x,\kappa_y;z)\right],
\label{EqS14}\\
\tilde{\bf H}^{\rm (m)}(\kappa_x,\kappa_y;z)&=&\frac{1}{\mu_r}\left\{k^2\tilde{{\bf \Pi}}^{\rm (m)}(\kappa_x,\kappa_y;z)-{\bf k}^{\pm}[{\bf k}^ {\pm}\cdot \tilde{{\bf \Pi}}^{\rm (m)}(\kappa_x,\kappa_y;z)]\right\},
\label{EqS15}
\end{eqnarray}
where $Z_0\equiv\sqrt{\mu_0/\varepsilon_0}$ is the impedance of free space. These expressions are readily obtained by the substitution $\nabla\to i{\bf k}^{\pm}$. Hence, cumbersome analysis involving the computation of gradients, divergences, curls, and Laplacian, is reduced to simple algebraic operations relying upon the product between the wavevector ${\bf k}^{\pm}$ and the spectral amplitudes $\tilde{{\bf \Pi}}^{\rm (e/m)}$.

\section{II. Angular spectrum representation of the electric dipole}

Below, for the sake of completeness and for convenience in the subsequent analysis, we will briefly sketch the derivation of the angular spectrum representation of an electric dipole. To do so, let us start by considering the corresponding complex-like vector potential associated to an oscillating electric dipole as given in Ref. \cite{Jacksonsm}: 
\begin{equation}
{\bf A}^{\rm ED}({\bf r})=\frac{\mu}{4\pi}\left[\int_{-\infty}^{+\infty}{{\bf J}({\bf r}')d{\bf r}'}\right] \frac{e^{ikr}}{r}.
\label{EqS16}
\end{equation}
After integration by parts and using the continuity equation for the electric charge, i.e., $\nabla\cdot {\bf J}+\partial_t\rho=0$, we can recast the latter expression in a more familiar form,
\begin{equation}
{\bf A}^{\rm ED}({\bf r})=-\frac{i\mu\omega}{4\pi}{\bf p}\frac{e^{ikr}}{r},
\label{EqS17}
\end{equation}
where ${\bf p}=\int_{-\infty}^{+\infty}{{\bf r}'\rho({\bf r}')d{\bf r}'}$ is the \textit{electric dipole moment}. This means that the electric current density can be actually expressed as ${\bf J}({\bf r})=-i\omega\delta^3({\bf r}-{\bf r}'){\bf p}$, and so the dipolar source is simply characterized by a constant vector ${\bf p}$. As pointed out above, to proceed with the angular spectrum representation, it is crucial to be certain that the vector potential given by Eq. \eqref{EqS17} really satisfies the Helmholtz equation:
\begin{equation}
\left(\nabla^2+k^2\right){\bf A}^{\rm ED}({\bf r})=0.
\label{EqS18}
\end{equation}
In this regard, it should be noted that, besides a constant prefactor, the vector potential only consists of a diverging spherical wave (i.e., $e^{ikr}/r$) which is in fact the scalar Green's function of the Helmholtz operator \cite{Mandelsm}, thereby satisfying in its own the Helmholtz equation everywhere except at the origin. Despite that singularity, we might formally get the angular spectrum of $e^{ikr}/r$ in each of the half-spaces $z>0$ and $z<0$ \cite{Mandelsm}. Hence, appealing to Weyl's identity given in Eq. \eqref{EqS1}, it is straightforward to show that 
\begin{equation}
{\bf A}^{\rm ED}({\bf r})=-\frac{i\mu \omega}{4\pi}{\bf p}\left[\frac{ik}{2\pi}\iint_{-\infty}^{+\infty}{\frac{1}{\kappa_z}e^{ik\left(\kappa_xx+\kappa_yy\pm \kappa_zz\right)}d\kappa_xd\kappa_y}\right].
\label{EqS19}
\end{equation}
By comparing the above expression with the general results given in Eqs. \eqref{EqS7} and \eqref{EqS8}, it is easy to see that the angular spectrum amplitude of the electric dipole is given by:
\begin{equation}
\tilde{{\bf A}}^{\rm ED}(\kappa_x,\kappa_y;z)=\tilde{{\bf A}}^{\rm ED}(\kappa_x,\kappa_y;0)e^{\pm ik\kappa_zz},
\label{EqS20}
\end{equation}
where
\begin{tcolorbox}[colback=blue!7,colframe=white!80!black]
\vspace{-0.25cm}
\begin{equation}
\tilde{{\bf A}}^{\rm ED}(\kappa_x,\kappa_y;0)=\frac{\mu c k^2}{8\pi^2 n}\frac{1}{\kappa_z}{\bf p}.
\label{EqS21}
\end{equation}
\end{tcolorbox}
\noindent A more detailed description of the above derivation, in particular the explicit calculations to obtain Weyl's identity analytically [Eq. \eqref{EqS1}], can be found in Ref. \cite{Mandelsm}.

\section{III. Angular spectrum representation of the electric quadrupole}

Building on the above scheme, in this section we will address the angular spectrum representation of the electric quadrupole. To this aim, we start once again from the corresponding vector potential as given in Ref. \cite{Jacksonsm}:
\begin{equation}
{\bf A}^{\rm EQ}({\bf r})=\frac{\mu\omega k}{8\pi}\left(\frac{1}{ikr}-1\right)\left[\int_{-\infty}^{+\infty}{{\bf r'}\left({\bf e}_r\cdot{\bf r}'\right)\rho({\bf r}')d{\bf r}'}\right] \frac{e^{ikr}}{r},
\label{EqS22}
\end{equation}
where ${\bf e}_r$ is the unit vector in the radial direction. Notice that this latter expression is given in terms of an integral involving the second moments of the charge density. Still, it can be readily simplified by using the following definitions:
\begin{equation}
\frac{{\bf Q}({\bf e}_r)}{3}\equiv\int_{-\infty}^{+\infty}{{\bf r}'\left({\bf e}_r\cdot{\bf r}'\right)\rho({\bf r}')d{\bf r}'},
\label{EqS23}
\end{equation}
where $Q_i=\sum_j{\mathcal{Q}_{ij}({\bf e}_r)_j}$, and 
\begin{equation}
\mathcal{Q}_{ij}=\int_{-\infty}^{+\infty}{\left(3r_ir_j-r^2\delta_{ij}\right)\rho({\bf r})d{\bf r}}
\label{EqS24}
\end{equation}
is the \textit{quadrupole moment tensor} characterizing the charge distribution. This is a rank-two tensor, usually represented by a $3\times3$ traceless, complex, and symmetric matrix, i.e., $\mathcal{Q}_{xx}+\mathcal{Q}_{yy}+\mathcal{Q}_{zz}=0$, and $\mathcal{Q}_{ij}=\mathcal{Q}_{ji}$, thus reducing the number of unknowns to $5$. Hence, the vector potential can be recast as
\begin{equation}
{\bf A}^{\rm EQ}({\bf r})=\frac{\mu\omega k}{24\pi}\left(\frac{1}{ikr}-1\right){\bf Q}({\bf e}_r)\frac{e^ {ikr}}{r}.
\label{EqS25}
\end{equation}
Looking at this expression more closely, one realizes that there are important discrepancies in comparison with the previous case of the electric dipole. On the one side, it should be noticed that ${\bf Q}({\bf e}_r)$ depends on the direction of observation. As a consequence, the scalar contribution of the electric quadrupole's vector potential is no longer a proper solution of the Helmholtz equation, and so we have to regard the whole vector expression of ${\bf A}^{\rm EQ}$ when addressing the angular spectrum representation. 

After verifying that Eq. \eqref{EqS25} does indeed satisfy the Helmholtz equation (it can be readily checked by means of a symbolic calculation software), we can then proceed to the computation of the angular spectrum amplitude for the electric quadrupole from the partial Fourier transform:
\begin{eqnarray}
\nonumber\tilde{{\bf A}}^{\rm EQ}(\kappa_x,\kappa_y;z)&=&\left(\frac{k}{2\pi}\right)^2\iint_{-\infty}^{+\infty}{{\bf A}^{\rm EQ}(x,y,z)e^{-ik\left(\kappa_xx+\kappa_yy\right)}dxdy}\\
&=&\frac{\mu\omega k^3}{96\pi^3}\iint_{-\infty}^{+\infty}{\left(\frac{1}{ik\sqrt{x^2+y^2+z^2}}-1\right){\bf Q}({\bf e}_r)\frac{e^{ik\sqrt{x^2+y^2+z^2}}}{\sqrt{x^2+y^2+z^2}}e^{-ik\left(\kappa_xx+\kappa_yy\right)}dxdy}.
\label{EqS26}
\end{eqnarray}
This integral can be easily solved making the following change of variables to cylindrical polar coordinates in both real and momentum space:
\begin{equation}
\begin{matrix*}[l]
x=R\cos{\theta},&\qquad&y=R\sin{\theta},&\qquad&z=z;\\
\kappa_x=\kappa_R\cos{\phi},&\qquad&\kappa_y=\kappa_R\sin{\phi},&\qquad&\kappa_z=\kappa_{z}.
\end{matrix*}
\label{EqS27}
\end{equation}
Therefore, Eq. \eqref{EqS26} reads as
\begin{equation}
\tilde{{\bf A}}^{\rm EQ}(\kappa_x,\kappa_y;z)=\frac{\mu\omega k^3}{96\pi^3}\int_{0}^{+\infty}{\int_0^{2\pi}{\left(\frac{1}{ik\sqrt{R^2+z^2}}-1\right){\bf Q}({\bf e}_r)\frac{e^{ik\sqrt{R^2+z^2}}}{\sqrt{R^2+z^2}}e^{-ik\kappa_R R\cos{\left(\theta-\phi\right)}}Rd\theta} dR},
\label{EqS28}
\end{equation}
where $R\equiv(x^2+y^2)^{1/2}$ and $\kappa_R\equiv(\kappa_x^2+\kappa_y^2)^{1/2}$ are both real and positive quantities. It is important to emphasize that ${\bf Q}({\bf e}_r)$ depends in magnitude and direction on the integration variables $R$ and $\theta$, so it cannot be taken outside the integral (as in the previous case for the electric dipole). Furthermore, this is indeed the term that conveys the ``quadrupolar character'' to the vector potential and it can be expressed as follows:
\begin{equation}
{\bf Q}\equiv \dvec{\mathcal{Q}}\cdot{\bf e}_r=Q_x{\bf e}_x+Q_y{\bf e}_y+Q_z{\bf e}_z,
\label{EqS29}
\end{equation}
where
\begin{equation}
Q_n=\displaystyle\frac{1}{\sqrt{R^2+z^2}}\left[R\left(\mathcal{Q}_{nx}\cos{\theta}+\mathcal{Q}_{ny}\sin{\theta}\right)+z\mathcal{Q}_{nz}\right],
\label{EqS30}
\end{equation}
with $n=\left\{x,y,z\right\}$. Then, with the aid of the following identities
\begin{equation}
\int_{0}^{2\pi}{e^{-i\alpha\cos(\theta-\beta)}d\theta}=2\pi J_0{\left(\alpha\right)},\qquad \qquad
\int_{0}^{2\pi}{e^{-i\alpha\cos{(\theta-\beta)}}\left\{\begin{matrix}
\cos{\theta}\\
\sin{\theta}
\end{matrix}\right\}d\theta}=-2\pi iJ_1{\left(\alpha\right)}\left\{\begin{matrix}
\cos{\beta}\\
\sin{\beta}
\end{matrix}\right\},
\label{EqS31}
\end{equation}
where $J_l(x)$ is the Bessel function of the first kind and order $l$, it follows at once that the integration with respect to the angle $\theta$ reduces to
\begin{equation}
\varOmega_n=\int_{0}^{2\pi}{\left[\dvec{\mathcal{Q}}\cdot{\bf e}_r\right]_ne^{-ik\kappa_RR\cos{\left(\theta-\phi\right)}}d\theta}=\frac{2\pi}{\sqrt{R^2+z^2}}\left[\mathcal{Q}_{nz}zJ_0(k\kappa_RR)-iRJ_1(k\kappa_RR)\left(\mathcal{Q}_{nx}\cos{\phi}+\mathcal{Q}_{ny}\sin{\phi}\right)\right].
\label{EqS32}
\end{equation}
It is worth remarking in this case the presence of Bessel functions of order $1$, in contrast with the case of the electric dipole where there are only Bessel functions of order $0$. In addition, it should be noted that the whole vectorial expression would become much simpler provided that we use the properties of the quadrupole moment tensor (traceless, symmetries, ...). From the latter result it can be observed that we can actually work with all vector components simultaneously by using an index $n=\left\{x,y,z\right\}$. In doing so it follows that
\begin{equation}
\tilde{A}^{\rm EQ}_n(\kappa_x,\kappa_y;z)\!=\!\frac{\mu\omega k^3}{48\pi^2}\!\!\int_{0}^{+\infty}{\!\!\left(\!\frac{1}{ik\sqrt{R^2+z^2}}\!-1\!\right)\! \frac{e^{ik\sqrt{R^2+z^2}}}{R^2+z^2}\!\left[\mathcal{Q}_{nz}zJ_0(k\kappa_RR)\!-\!iRJ_1(k\kappa_RR)\!\left(\mathcal{Q}_{nx}\cos{\phi}\!+\!\mathcal{Q}_{ny}\sin{\phi}\right)\right]\!RdR}.
\label{EqS33}
\end{equation}
Now we only have to perform the integration over the radial variable. In order to do that analytically, likewise as for the electric dipole, we ought to take the limit $z\to 0$. Nonetheless, in this case, it is important to realize that taking the limit to the plane $z=0$ may lead to misleading outcomes wherein the $z$-dependent contribution of the quadrupole moment tensor would seem to be dismissed. Thus, to overcome this issue we should carry out the integration separately for the $xy$ and $z$ components of the quadrupole moment (i.e., $\tilde{A}_n^{\rm EQ}=\tilde{A}_{n,xy}^{\rm EQ}+\tilde{A}_{n,z}^{\rm EQ}$), where:
\begin{eqnarray}
\nonumber\tilde{A}_{n,xy}^{\rm EQ}(\kappa_x,\kappa_y;z)&=&\frac{i\mu\omega k^3}{48\pi^2}\int_{0}^{+\infty}{\left(1-\frac{1}{ik\sqrt{R^2+z^2}}\right)\frac{e^{ik\sqrt{R^2+z^2}}}{R^2+z^2}R^2J_1(k\kappa_RR)\left(\mathcal{Q}_{nx}\cos{\phi}+\mathcal{Q}_{ny}\sin{\phi}\right)dR}\\
\nonumber&\xrightarrow[z\to 0]{}&\frac{i\mu\omega k^3}{48\pi^2}\int_{0}^{+\infty}{\left(1-\frac{1}{ikR}\right)e^{ikR}J_1(k\kappa_RR)\left(\mathcal{Q}_{nx}\cos{\phi}+\mathcal{Q}_{ny}\sin{\phi}\right)dR}\\
&=&-\frac{i\mu\omega k^2}{48\pi^2}\frac{\kappa_R}{\kappa_z}\left(\mathcal{Q}_{nx}\cos{\phi}+\mathcal{Q}_{ny}\sin{\phi}\right)=\tilde{A}_{n,xy}^{\rm EQ}(\kappa_x,\kappa_y;0).
\label{EqS34}
\end{eqnarray}
As anticipated earlier, special care must be taken with the integral involving the $z$-dependent contribution of $\dvec{\mathcal{Q}}$. Then, we make use of integration by parts:
\begin{eqnarray}
\nonumber\tilde{A}^{\rm EQ}_{n,z}(\kappa_x,\kappa_y;z)&=&\frac{\mu\omega k^3}{48\pi^2}\int_{0}^{+\infty}{\left(\frac{1}{ik\sqrt{R^2+z^2}}-1\right)\frac{e^{ik\sqrt{R^2+z^2}}}{R^2+z^2}RzJ_0(k\kappa_RR)\mathcal{Q}_{nz}dR}\\
\nonumber&=&\left\{\begin{matrix*}[l]
u=J_0(k\kappa_RR)&\quad \Longrightarrow \quad & du=-k\kappa_RJ_1(k\kappa_RR)dR\\
dv=\displaystyle\left(\frac{1}{ik\sqrt{R^2+z^2}}-1\right)\frac{Rz}{R^2+z^2}e^{ik\sqrt{R^2+z^2}}dR&\quad \Longrightarrow \quad & v=\displaystyle\frac{ize^{ik\sqrt{R^2+z^2}}}{k\sqrt{R^2+z^2}}
\end{matrix*}\right\}\\
\nonumber&=&\frac{\mu\omega k^3}{48\pi^2}\mathcal{Q}_{nz}\left[\left.\frac{iz}{k\sqrt{R^2+z^2}}J_0(k\kappa_RR)e^{ik\sqrt{R^2+z^2}}\right|_{R=0}^{R\to\infty}+\int_{0}^{+\infty}{\frac{iz\kappa_R}{\sqrt{R^2+z^2}}J_1(k\kappa_RR)e^{ik\sqrt{R^2+z^2}}dR}\right]\\
\nonumber&=&\frac{\mu\omega k^3}{48\pi^2}\mathcal{Q}_{nz}\left[-\frac{i}{k}\frac{z}{|z|}e^{ik|z|}+\int_{0}^{+\infty}{\frac{iz\kappa_R}{\sqrt{R^2+z^2}}J_1(k\kappa_RR)e^{ik\sqrt{R^2+z^2}}dR}\right]\\
&\xrightarrow[z\to 0]{}&\mp\frac{i\mu\omega k^2}{48\pi^2}\mathcal{Q}_{nz}=\tilde{A}_{n,z}^{\rm EQ}(\kappa_x,\kappa_y;0).
\label{EqS35}
\end{eqnarray}
Hence, gathering together the above calculations, the spectral amplitude of the electric quadrupole reads
\begin{eqnarray}
\nonumber\tilde{A}_n^{\rm EQ}(\kappa_x,\kappa_y;0)&=&-\frac{i\mu\omega k^2}{48\pi^2}\frac{1}{\kappa_z}\left[\kappa_R\left(\mathcal{Q}_{nx}\cos{\phi}+\mathcal{Q}_{ny}\sin{\phi}\right)\pm k_z \mathcal{Q}_{nz}\right]\\
&=&-\frac{i\mu\omega k^2}{48\pi^2}\frac{1}{\kappa_z}\left[\kappa_x\mathcal{Q}_{nx}+\kappa_y\mathcal{Q}_{ny}\pm \kappa_z \mathcal{Q}_{nz}\right].
\label{EqS36}
\end{eqnarray}
Since ${\bf k}^{\pm}=k(\kappa_x,\kappa_y,\pm \kappa_z)$ (where signs $+$ and $-$ correspond to $z>0$ and $z<0$, respectively), the above result can be expressed in a closed form:
\begin{tcolorbox}[colback=blue!7,colframe=white!80!black]
\vspace{-0.25cm}
\begin{equation}
\tilde{{\bf A}}^{\rm EQ}(\kappa_x,\kappa_y;0)=\frac{\mu c k^2}{8\pi^2n}\frac{1}{\kappa_z}\frac{(-i)\left[\dvec{\mathcal{Q}}\cdot{\bf k}^{\pm}\right]}{6}.
\label{EqS37}
\end{equation}
\end{tcolorbox}
\noindent Notice that we have deliberately written down the latter expression in such a way that is straightforward to compare it with the angular spectrum amplitude of the electric dipole [Eq. \eqref{EqS21}]. In this respect, it is worth pointing out that we can attain the spectral amplitude of the electric quadrupole from that of the electric dipole by simply performing the substitution ${\bf p}\to (-i)\left[\dvec{\mathcal{Q}}\cdot{\bf k}^{\pm}\right]/6$. This additional ${\bf k}$ dependence on the spectral amplitude has important consequences regarding the spatial range where the evanescent modes have a significant contribution. 

\section{IV. Spectral amplitudes for the electric and magnetic fields of the electric dipole and quadrupole}

Until now, we have been mainly focused on the angular spectra of the vector potentials. However, we are typically more interested in knowing the angular spectrum representation of the EM fields themselves. As has been already pointed out above, we can write the complex fields ${\bf E}({\bf r})$ and ${\bf H}({\bf r})$ from the standard vector potential ${\bf A}({\bf r})$ by means of the following relations:
\begin{eqnarray}
{\bf B}({\bf r})=\nabla\times{\bf A}({\bf r})&\qquad\Longrightarrow\qquad&{\bf H}({\bf r})=\frac{1}{\mu}[\nabla\times{\bf A}({\bf r})];
\label{EqS38}\\
\nabla\times{\bf H}({\bf r})=-i\omega\varepsilon{\bf E}({\bf r})&\qquad\Longrightarrow\qquad& {\bf 
E}({\bf r})=\frac{ic}{kn}\left[\nabla\left(\nabla\cdot{\bf A}({\bf r})\right)-\nabla^2{\bf A}({\bf r})\right].
\label{EqS39}
\end{eqnarray}
It should be noted that, for the moment, we skip the usage of Hertz potentials (they will turn out to be relevant below when dealing with EM multipole fields). Taking advantage of the angular spectrum formalism, one can easily represent the complex field amplitudes of both the electric dipole and quadrupole in a compact notation as
\begin{eqnarray}
\tilde{\bf E}^{\rm ED}(\kappa_x,\kappa_y;0)&=&\frac{ic}{kn}[k^2\tilde{\bf A}^{\rm ED}_0-{\bf k}^{\pm}({\bf k}^{\pm}\cdot\tilde{\bf A}^{\rm ED}_0)]=\frac{ik}{8\pi^2\varepsilon}\frac{1}{\kappa_z}\left\{k^2{\bf p}-{\bf k}^{\pm}({\bf k}^{\pm}\cdot{\bf p})\right\},
\label{EqS40}\\
\tilde{\bf H}^{\rm ED}(\kappa_x,\kappa_y;0)&=&\frac{i}{\mu}[{\bf k}^\pm\times \tilde{\bf A}^{\rm ED}_0]=\frac{i\omega k}{8\pi^2}\frac{1}{\kappa_z}\left[{\bf k}^{\pm}\times{\bf p}\right];
\label{EqS41}\\
\tilde{\bf E}^{\rm EQ}(\kappa_x,\kappa_y;0)&=&\frac{ic}{kn}[k^2\tilde{\bf A}^{\rm EQ}_0-{\bf k}^{\pm}({\bf k}^{\pm}\cdot\tilde{\bf A}^{\rm EQ}_0)]=\frac{k}{48\pi^2\varepsilon}\frac{1}{\kappa_z}\left\{k^2(\dvec{\mathcal{Q}}\cdot{\bf k}^{\pm})-{\bf k}^{\pm}[{\bf k}^{\pm}\cdot(\dvec{\mathcal{Q}}\cdot{\bf k}^{\pm})]\right\},
\label{EqS42}\\
\tilde{\bf H}^{\rm EQ}(\kappa_x,\kappa_y;0)&=&\frac{i}{\mu}[{\bf k}^\pm\times \tilde{\bf A}^{\rm EQ}_0]=\frac{\omega k}{48\pi^2}\frac{1}{\kappa_z}[{\bf k}^{\pm}\times(\dvec{\mathcal{Q}}\cdot{\bf k}^{\pm})],
\label{EqS43}
\end{eqnarray}
where, so as to avoid cumbersome notation, $\tilde{\bf A}^{\rm ED/EQ}_0\equiv \tilde{\bf A}^{\rm ED/EQ}(\kappa_x,\kappa_y;z=0)$. Further simplifications can be made by introducing the polarization vector basis \cite{Picardi2017sm}:
\begin{eqnarray}
{\bf e}_s&\equiv&\frac{{\bf e}_z\times{\bf k}^{\pm}}{\sqrt{({\bf e}_z\times{\bf k}^{\pm})\cdot({\bf e}_z\times{\bf k}^{\pm})}}=\frac{(-\kappa_y,\kappa_x,0)}{\kappa_R},
\label{EqS44}\\
{\bf e}_p^{\pm}&\equiv&{\bf e}_s\times{\bf k}^{\pm}/k=\frac{(\pm\kappa_x\kappa_z,\pm \kappa_y\kappa_z,-(\kappa_x^2+\kappa_y^2))}{\kappa_R},
\label{EqS45}
\end{eqnarray}
where the subscripts indicate the $s$ and $p$ polarizations \cite{footnote1sm}, and ${\bf e}_z$ is the unit vector perpendicular to the plane in which we perform the angular spectrum representation. Notice that in the propagating lossless case (i.e., when ${\bf k}^{\pm}$ is real), these two polarization vectors correspond to the usual unit vectors in spherical coordinates ${\bf e}_\varphi$ and ${\bf e}_\theta$, respectively, and form, together with ${\bf k}^{\pm}/k$ an orthonormal basis. However, even when we consider a complex ${\bf k}^{\pm}$, the triad $\left\{{\bf e}_s, {\bf e}_p^{\pm}, {\bf k}^{\pm}/k\right\}$ remains orthonormal as long as we define the inner product as the dot product with no complex conjugation: ${\bf e}_s\cdot{\bf e}_s={\bf e}_p^\pm\cdot{\bf e}_p^{\pm}=({\bf k}^{\pm}\cdot{\bf k}^{\pm})/k^2=1$, and ${\bf e}_s\cdot{\bf e}_p^{\pm}={\bf k}^{\pm}\cdot{\bf e}_s={\bf k}^{\pm}\cdot{\bf e}_p^{\pm}=0$. Moreover, it can be shown that 
\begin{equation}
{\bf k}^{\pm}\times{\bf e}_s=-k{\bf e}_p^{\pm},\qquad\qquad\qquad \qquad {\bf k}^{\pm}\times{\bf e}_p^{\pm}=k{\bf e}_s.
\label{EqS46}
\end{equation}
Therefore, taking into account the above relations together with the scalar triple product identity, ${\bf A}\cdot\left({\bf B}\times{\bf C}\right)={\bf B}\cdot\left({\bf C}\times{\bf A}\right)={\bf C}\cdot\left({\bf A}\times{\bf B}\right)$, we can recast the EM fields as the projections along the polarization vectors:
\begin{tcolorbox}[colback=blue!7,colframe=white!80!black]
\vspace{-0.25cm}
\begin{eqnarray}
\tilde{\bf E}^{\rm ED}(\kappa_x,\kappa_y;0)&=&\frac{ik^3}{8\pi^2\varepsilon}\frac{1}{\kappa_z}\left\{[{\bf e}_s\cdot{\bf p}]{\bf e}_s+[{\bf e}_p^\pm\cdot{\bf p}]{\bf e}_p^\pm\right\},
\label{EqS47}\\
\tilde{\bf H}^{\rm ED}(\kappa_x,\kappa_y;0)&=&\frac{i k^3}{8\pi^2 \varepsilon}\sqrt{\frac{\varepsilon}{\mu}}\frac{1}{\kappa_z}\left\{[{\bf e}_p^\pm\cdot {\bf p}]{\bf e}_s-[{\bf e}_s\cdot{\bf p}]{\bf e}_p^{\pm}\right\};
\label{EqS48}\\
\tilde{\bf E}^{\rm EQ}(\kappa_x,\kappa_y;0)&=&\frac{k^3}{48\pi^2\varepsilon}\frac{1}{\kappa_z}\left\{[{\bf e}_s\cdot(\dvec{\mathcal{Q}}\cdot{\bf k}^{\pm})]{\bf e}_s+[{\bf e}_p^\pm\cdot(\dvec{\mathcal{Q}}\cdot{\bf k}^{\pm})]{\bf e}_p^\pm\right\},
\label{EqS49}\\
\tilde{\bf H}^{\rm EQ}(\kappa_x,\kappa_y;0)&=&\frac{k^3}{48\pi^2\varepsilon}\sqrt{\frac{\varepsilon}{\mu}}\frac{1}{\kappa_z}\left\{[{\bf e}_p^\pm\cdot (\dvec{\mathcal{Q}}\cdot{\bf k}^{\pm})]{\bf e}_s-[{\bf e}_s\cdot(\dvec{\mathcal{Q}}\cdot{\bf k}^{\pm})]{\bf e}_p^{\pm}\right\}.
\label{EqS50}
\end{eqnarray}
\end{tcolorbox}

Summarizing all the above, so far we have presented the Weyl's identity, the expression for the angular spectrum representation and the Hertz potentials. We have also sketched the angular spectrum for the particular case of the electric dipole and afterwards, we have performed the corresponding full derivation for the electric quadrupole (notice that we have omitted the magnetic dipole as yet; see appendix for details). This has enabled us to get the spectral representation of the EM fields in a very simple manner in terms of the $s$ and $p$ transverse unit vectors. Even so, this approach essentially relies on the ability to determine firstly the vector potential from the following expansion \cite{Jacksonsm}:
\begin{equation}
{\bf A}({\bf r})=\frac{\mu}{4\pi}\int_{-\infty}^{+\infty}{{\bf J}({\bf r}')\frac{e^{ik|{\bf r}-{\bf r}'|}}{|{\bf r}-{\bf r}'|}d{\bf r}'}.
\label{EqS51}
\end{equation}
Moreover we must be able to identify either the electric or the magnetic nature of the corresponding optical source. Thus, to find a formulation as general as possible, including the angular spectrum representation of any multipole field, we would have to go through higher-order terms in the previous expansion of the standard vector potential one by one. However, as pointed out by Jackson in Ref. \cite{Jacksonsm}, this is only feasible for the lowest orders, i.e., just for the electric and magnetic dipole, or even for the electric quadrupole at best:
\begin{quote}
\textit{``[\ldots ] The labor involved in manipulating higher terms in expansion of the vector potential becomes increasingly prohibitive as the expansion is extended beyond the electric quadrupole terms. Another disadvantage of the present approach is that physically distinct fields such as those of the magnetic dipole and the electric quadrupole must be disentangled from the separate terms [\ldots]''}
\end{quote}
Owing to these considerations, in the next section we shall draw on the concepts previously introduced to provide a more suitable and systematic formulation for multipolar sources of arbitrary order.

\section{V. Angular spectrum representation of the electromagnetic multipole fields}

In this section we aim to find the angular spectrum representation of the general electric and magnetic multipole fields. Leaving aside the detailed procedure to obtain their explicit expressions, which can be found elsewhere (see, e.g., Refs. \cite{Jacksonsm,Novotnysm}), the electric (e) and magnetic (m) multipole fields of arbitrary order can be written as follows:
\begin{eqnarray}
{\bf H}_{l,m}^{\rm (e)}({\bf r})&=&\frac{f_l(kr)}{\sqrt{l(l+1)}}{\bf L}Y_{l,m}(\Omega),
\label{EqS52}\\
{\bf E}_{l,m}^{\rm (e)}({\bf r})&=&\frac{i}{\varepsilon\omega}[\nabla\times{\bf H}_{l,m}^{\rm (e)}({\bf r})],
\label{EqS53}\\
{\bf E}_{l,m}^{\rm (m)}({\bf r})&=&\frac{\mu c}{n}\frac{f_l(kr)}{\sqrt{l(l+1)}}{\bf L}Y_{l,m}(\Omega),
\label{EqS54}\\
{\bf H}_{l,m}^{\rm (m)}({\bf r})&=&-\frac{i}{\mu \omega}[\nabla\times{\bf E}_{l,m}^{\rm (m)}({\bf r})],
\label{EqS55}
\end{eqnarray}
where $Y_{l,m}(\Omega)$ are the spherical harmonics of order $(l,m)$, $\Omega\equiv(\theta,\varphi)$ stands for the standard angular coordinates (i.e., polar and azimuthal angles, respectively), $f_l(kr)\equiv\left\{j_l(kr),y_l(kr)\right\}$ are the $l$-dependent spherical Bessel functions of first and second kind, and ${\bf L}\equiv-i\left({\bf r}\times \nabla\right)$ is the orbital angular momentum operator. Since there are two independent radial dependent solutions [$j_l(kr)$ and $y_l(kr)$], any linear combination will also be a solution, and specifically the spherical Hankel functions $h_l^{(\pm)}(kr)=j_l(kr)\pm iy_{l}(kr)$ are used to represent propagating spherical waves.

Rather than dealing with the EM fields themselves, it is convenient to resort to the vector potentials. In this case, however, instead of considering the standard vector potentials, we will use a formulation based on the Hertz potentials to treat the electric and magnetic multipole fields on an equal footing. Under this representation, the first step will consist in determining an explicit expression for both the electric and magnetic Hertz potentials associated to the multipole fields of arbitrary order. This can be easily done just by looking at the Eqs. \eqref{EqS10} and \eqref{EqS11}, which allow us to relate the magnetic and the electric fields to the electric and magnetic Hertz potentials, respectively:
\begin{eqnarray}
{\bf H}_{l,m}^{\rm (e)}({\bf r})&=&-i\omega[\nabla\times{\bf \Pi}_{l,m}^{\rm (e)}({\bf r})],
\label{EqS56}\\
{\bf E}_{l,m}^{\rm (m)}({\bf r})&=&i\mu_0\omega[\nabla\times{\bf \Pi}_{l,m}^{\rm (m)}({\bf r})].
\label{EqS57}
\end{eqnarray}
Taking also into account the definitions given in Eqs. \eqref{EqS52} and \eqref{EqS54} it follows that
\begin{eqnarray}
{\bf H}_{l,m}^{\rm (e)}({\bf r})&=&-i\omega[\nabla\times{\bf \Pi}_{l,m}^{\rm (e)}({\bf r})]=\frac{h_l(kr)}{\sqrt{l(l+1)}}{\bf L}Y_{l,m}(\Omega)=\frac{h_l(kr)}{i\sqrt{l(l+1)}}\left[{\bf r}\times\nabla Y_{l,m}(\Omega)\right],
\label{EqS58}\\
{\bf E}_{l,m}^{\rm (m)}({\bf r})&=&i\mu_0\omega[\nabla\times{\bf \Pi}_{l,m}^{\rm (m)}({\bf r})]=\frac{\mu c}{n}\frac{h_l(kr)}{\sqrt{l(l+1)}}{\bf L}Y_{l,m}(\Omega)=\frac{\mu c}{n}\frac{h_l(kr)}{i\sqrt{l(l+1)}}\left[{\bf r}\times\nabla Y_{l,m}(\Omega)\right],
\label{EqS59}
\end{eqnarray}
where we have particularized the spherical Hankel functions as the radial functions. With the help of the identity ${\bf v}\times\nabla f=f\nabla\times{\bf v}-\nabla\times\left(f{\bf v}\right)$, the latter expressions can be recast as
\begin{eqnarray}
{\bf H}_{l,m}^{\rm (e)}({\bf r})&=&\frac{h_l(kr)}{i\sqrt{l(l+1)}}\left[Y_{l,m}(\Omega)\nabla\times{\bf r}-\nabla\times\left(Y_{l,m}(\Omega){\bf r}\right)\right]=i\frac{h_l(kr)}{\sqrt{l(l+1)}}\left[\nabla\times\left(Y_{l,m}(\Omega){\bf r}\right)\right],
\label{EqS60}\\
{\bf E}_{l,m}^{\rm (m)}({\bf r})&=&\frac{\mu c}{n}\frac{h_l(kr)}{i\sqrt{l(l+1)}}[Y_{l,m}(\Omega)\nabla\times{\bf r}-\nabla\times\left(Y_{l,m}(\Omega){\bf r}\right)]=\frac{i\mu c}{n}\frac{h_l(kr)}{\sqrt{l(l+1)}}[\nabla\times\left(Y_{l,m}(\Omega){\bf r}\right)],
\label{EqS61}
\end{eqnarray}
where we used the fact that $\nabla\times{\bf r}=0$. It is important to realize that, since the only vector component is the radial one, we can judiciously insert the $r$-dependent prefactor inside the curl operator:
\begin{eqnarray}
{\bf H}_{l,m}^{\rm (e)}({\bf r})&=&\nabla\times\left(\frac{i}{\sqrt{l(l+1)}}rh_l(kr)Y_{l,m}(\Omega){\bf e}_r\right),
\label{EqS62}\\
{\bf E}_{l,m}^{\rm (m)}({\bf r})&=&\nabla\times\left(\frac{i\mu c}{n\sqrt{l(l+1)}}rh_l(kr)Y_{l,m}(\Omega){\bf e}_r\right).
\label{EqS63}
\end{eqnarray}
By comparing these results with the definitions of the EM multipole fields in terms of the Hertz potentials [see Eqs. \eqref{EqS56} and \eqref{EqS57}], it is straightforward to show that a first attempt at finding potentials for arbitrary electric and magnetic multipole fields leads to
\begin{eqnarray}
{\bf \Pi}^{\rm rad(e)}_{l,m}({\bf r})&=&-\frac{1}{\omega\sqrt{l(l+1)}}rh_l(kr)Y_{l,m}(\Omega){\bf e}_r,
\label{EqS64}\\
{\bf \Pi}^{\rm rad(m)}_{l,m}({\bf r})&=&\sqrt{\frac{\mu_r}{\varepsilon_r}}\frac{ c}{\omega\sqrt{l(l+1)}}rh_l(kr)Y_{l,m}(\Omega){\bf e}_r.
\label{EqS65}
\end{eqnarray}
However, despite their simplicity, it can be demonstrated that the above expressions fail when we attempt to address the angular spectrum representation. This is because they do not satisfy the Helmholtz wave equation: $\left(\nabla^2+k^2\right){\bf \Pi}^{\rm rad(e/m)}_{l,m}\neq 0$. Therefore, referring to ${\bf \Pi}^{\rm rad(e/m)}_{l,m}$ as the Hertz potentials is somehow an abuse of terminology. Nonetheless, we can overcome this issue just by adding the corresponding gradient of the electric- and magnetic-like gauge functions:
\begin{eqnarray}
\chi_{l,m}^{\rm (e)}({\bf r})&=&-\frac{1}{\omega\sqrt{l(l+1)}}\frac{r}{k}h_{l-1}(kr)Y_{l,m}(\Omega),
\label{EqS66}\\
\chi_{l,m}^{\rm (m)}({\bf r})&=&\sqrt{\frac{\mu_r}{\varepsilon_r}}\frac{c}{\omega\sqrt{l(l+1)}}\frac{r}{k}h_{l-1}(kr)Y_{l,m}(\Omega).
\label{EqS67}
\end{eqnarray} 
Hence, the vector potentials now read as
\begin{eqnarray}
{\bf \Pi}^{\rm (e)}_{l,m}({\bf r})&=&{\bf \Pi}^{\rm rad(e)}_{l,m}({\bf r})+\nabla \chi^{\rm (e)}_{l,m}({\bf r})=-\frac{1}{\omega k\sqrt{l(l+1)}}\left\{krh_l(kr)Y_{l,m}(\Omega){\bf e}_r+\nabla\left[rh_{l-1}(kr)Y_{l,m}(\Omega)\right]\right\},
\label{EqS68}\\
{\bf \Pi}^{\rm (m)}_{l,m}({\bf r})&=&{\bf \Pi}_{l,m}^{\rm rad(m)}({\bf r})+\nabla\chi_{l,m}^{\rm (m)}({\bf r})=\sqrt{\frac{\mu_r}{\varepsilon_r}}\frac{c}{\omega k\sqrt{l(l+1)}}\left\{krh_{l}(kr)Y_{l,m}(\Omega){\bf e}_r+\nabla\left[rh_{l-1}(kr)Y_{l,m}(\Omega)\right]\right\}.
\label{EqS69}
\end{eqnarray}
These results can be further simplified by expanding the gradient of the gauge function:
\begin{eqnarray}
\nonumber\nabla\left[rh_{l-1}(kr)Y_{l,m}(\Omega)\right]&=&\left[h_{l-1}(kr)+r\partial_rh_{l-1}(kr)\right]Y_{l,m}(\Omega){\bf e}_r+rh_{l-1}(kr)\nabla Y_{l,m}(\Omega)\\
&=&\left[krh_{l-2}-(l-1)h_{l-1}(kr)\right]Y_{l,m}(\Omega){\bf e}_r+rh_{l-1}(kr)\nabla Y_{l,m}(\Omega).
\label{EqS70}
\end{eqnarray}
By means of the recurrence relation $krh_{l-2}-(l-1)h_{l-1}+krh_l=lh_{l-1}$ we then arrive at
\begin{eqnarray}
{\bf \Pi}^{\rm (e)}_{l,m}({\bf r})&=&-\frac{1}{\omega k\sqrt{l(l+1)}}h_{l-1}(kr)\left[lY_{l,m}(\Omega){\bf e}_r+r\nabla Y_{l,m}(\Omega)\right],
\label{EqS71}\\
{\bf \Pi}^{\rm (m)}_{l,m}({\bf r})&=&\sqrt{\frac{\mu_r}{\varepsilon_r}}\frac{c}{\omega k\sqrt{l(l+1)}}h_{l-1}(kr)\left[lY_{l,m}(\Omega){\bf e}_r+r\nabla Y_{l,m}(\Omega)\right].
\label{EqS72}
\end{eqnarray}
Furthermore, aiming to get a closed expression, we make use of the explicit definition of the spherical harmonics \cite{Jacksonsm}:
\begin{equation}
Y_{l,m}(\Omega)\equiv \sqrt{\frac{2l+1}{4\pi}\frac{\left(l-m\right)!}{\left(l+m\right)!}}P_l^m{\left(\cos\theta\right)}e^{im\varphi}=C_{l,m}P_l^m{\left(\cos\theta\right)}e^{im\varphi},
\label{EqS73}
\end{equation}
where $P_l^m{\left(\cos\theta\right)}$ are the associated Legendre polynomials of degree $l$ and order $m$, and $C_{l,m}$ is used as a shorthand notation for the prefactor. In addition, it should be noted that $\displaystyle Y_{l,-m}=(-1)^{m}Y_{l,m}^*$, with the asterisk denoting complex conjugation. Then, it can be shown that
\begin{equation}
lY_{l,m}(\Omega){\bf e}_r+r\nabla Y_{l,m}(\Omega)=C_{l,m}\left[lP_l^m(\cos{\theta}){\bf e}_r+\partial_\theta P_l^m({\cos{\theta}}){\bf e}_\theta+\frac{im}{\sin{\theta}}P_l^m(\cos{\theta}){\bf e}_\varphi\right]e^{im\varphi},
\label{EqS74}
\end{equation}
where ${\bf e}_r\equiv\left(\sin{\theta}\cos{\varphi},\sin{\theta}\sin{\varphi},\cos{\theta}\right)$, ${\bf e}_\theta\equiv\left(\cos{\theta}\cos{\varphi},\cos{\theta}\sin{\varphi},-\sin{\theta}\right)$, and ${\bf e}_\varphi\equiv\left(-\sin{\varphi},\cos{\varphi},0\right)$, are the spherical unit vectors in Cartesian basis. Finally, after some straightforward but lengthy manipulations we arrive at a closed expression for a valid Hertz potential of any electric and magnetic multipolar source of arbitrary order:
\begin{tcolorbox}[colback=blue!7,colframe=white!80!black]
\vspace{-0.25cm}
\begin{equation}
{\bf \Pi}^{\rm (e)}_{l,m}({\bf r})=-\frac{C_{l,m}h_{l-1}(kr)}{\omega k\sqrt{l(l+1)}}\left[\alpha_{l,m}(\theta)\cos{\varphi}-m\beta_{l,m}(\theta)e^{i\varphi},\alpha_{l,m}(\theta)\sin{\varphi}+im\beta_{l,m}(\theta)e^{i\varphi},\gamma_{l,m}(\theta)\right]e^{im\varphi},
\label{EqS75}
\end{equation}
and
\begin{equation}
{\bf \Pi}^{\rm (m)}_{l,m}({\bf r})=-c\sqrt{\frac{\mu_r}{\varepsilon_r}}{\bf \Pi}^{\rm (e)}_{l,m}({\bf r}),
\label{EqS76}
\end{equation}
with
\begin{equation}
\begin{aligned}
\alpha_{l,m}(\theta)&=\frac{1}{\sin{\theta}}\left[(l+m)\left(P_l^m(\cos{\theta})-\cos{\theta}P_{l-1}^m(\cos{\theta})\right)\right],\\ \beta_{l,m}(\theta)&=\frac{1}{\sin{\theta}}P_l^m(\cos{\theta}),\\
\gamma_{l,m}(\theta)&=(l+m)P_{l-1}^m(\cos{\theta}).
\end{aligned}
\label{EqS77}
\end{equation}
\end{tcolorbox}
\noindent As we can see, using spherical coordinates, the radial, polar and azimuthal dependences appear to be separated, thereby simplifying the subsequent calculations of the partial Fourier transform. In addition, it should also be noted that these expressions can be greatly simplified when we restrict ourselves to the plane $z=0$, i.e., for $\theta=\pi/2$. In fact, in this particular case the $\theta$-dependent functions become,
\begin{equation}
\begin{aligned}
  \alpha_{l,m}(\pi/2)&=\left(l+m\right)P_l^m(0)=\left(\frac{l+m}{l+m+1}\right)\gamma_{l+1,m}(\pi/2),\\
  \beta_{l,m}(\pi/2)&=P_l^m(0)=\left(\frac{1}{l+m+1}\right)\gamma_{l+1,m}(\pi/2),\\
  \gamma_{l,m}(\pi/2)&=\left(l+m\right)P_{l-1}^m(0),
\end{aligned}
\label{EqS78}
\end{equation}
where
\begin{equation}
P_l^m(0)=\left\{\begin{matrix*}[l]
\displaystyle (-1)^{l+\frac{l-m}{2}}\frac{\left(l-m+2m\right)!}{2^l\left(\frac{l-m}{2}\right)!\left(\frac{l+m}{2}\right)!}, & \quad \text{for} \quad & l+m \quad \text{even};\\\\
0& \quad \text{for} \quad & l+m \quad \text{odd}.
\end{matrix*}\right.
\label{EqS79}
\end{equation}

Having an explicit expression for both the electric and magnetic Hertz vector potential [Eqs. \eqref{EqS75} and \eqref{EqS76}], we now proceed with finding their angular spectrum at the plane $z=0$. Indeed, since ${\bf \Pi}^{\rm (e)}_{l,m}$ (and consequently ${\bf \Pi}^{\rm (m)}_{l,m}$) satisfies the Helmholtz equation, the spectral amplitudes of the electric multipole fields are given by the partial Fourier transform of Eq. \eqref{EqS75}:
\begin{eqnarray}
\!\!\!\!\!\!\!\!\!\!\!\!\nonumber\tilde{{\bf \Pi}}^{\rm (e)}_{l,m}\!(\kappa_x,\kappa_y;z)
\!&=&\!\left(\frac{k}{2\pi}\right)^2\!\!\iint_{-\infty}^{+\infty}{\!\!{\bf \Pi}^{\rm (e)}_{l,m}(x,y,z)e^{-ik\left(\kappa_{x}x+\kappa_{y}y\right)}dxdy}\\
\!&=&\!\frac{-k}{4\pi^2 \omega }\frac{C_{l,m}}{\sqrt{l(l+1)}}\int_{0}^{+\infty}{\!\!\!\int_{0}^{2\pi}{\!\!h_{l-1}(kr)\!\left\{\mathcal{X}^{\rm (e)}_{l,m}(\Omega),\mathcal{Y}^{\rm (e)}_{l,m}(\Omega),\mathcal{Z}^{\rm (e)}_{l,m}(\Omega)\right\}\!e^{-ik\kappa_Rr\sin{\theta}\cos{\left(\varphi-\phi\right)}}r\sin{\theta}d\varphi}dr},
\label{EqS80}
\end{eqnarray}
where 
\begin{eqnarray}
\mathcal{X}_{l,m}^{\rm (e)}(\Omega)&=&\left[\alpha_{l,m}(\theta)\cos{\varphi}-m\beta_{l,m}(\theta)e^{i\varphi}\right]e^{im\varphi},
\label{EqS81}\\
\mathcal{Y}_{l,m}^{\rm (e)}(\Omega)&=&\left[\alpha_{l,m}(\theta)\sin{\varphi}+im\beta_{l,m}(\theta)e^{i\varphi}\right]e^{im\varphi},
\label{EqS82}\\
\mathcal{Z}_{l,m}^{\rm (e)}(\Omega)&=&\gamma_{l,m}(\theta)e^{im\varphi}.
\label{EqS83}
\end{eqnarray}
To perform the integration along the angle $\varphi$ we will consider each of the vector components separately:
\begin{eqnarray*}
\int_{0}^{2\pi}{\!\!\mathcal{X}_{l,m}^{\rm (e)}e^{-ik\kappa_Rr\sin{\theta}\cos{\left(\varphi-\phi\right)}}d\varphi}&=&\int_{0}^{2\pi}{\!\left[\alpha_{l,m}\cos{\varphi}e^{i\left(m\varphi-k\kappa_Rr\sin{\theta}\cos{\left(\varphi-\phi\right)}\right)}\!-\!m\beta_{l,m}e^{i\left[\left(m+1\right)\varphi-k\kappa_Rr\sin{\theta}\cos{\left(\varphi-\phi\right)}\right]}\right]d\varphi}\\
&=&\pi(-i)^{m+1} \left[\left(\alpha_{l,m}-2m\beta_{l,m}\right) J_{m+1}(k\kappa_Rr\sin{\theta})e^{i\phi}-\alpha_{l,m} J_{m-1}(k\kappa_Rr\sin{\theta})e^{-i\phi}\right]e^{im\phi},\\\\
\int_{0}^{2\pi}{\!\!\mathcal{Y}_{l,m}^{\rm (e)}e^{-ik\kappa_Rr\sin{\theta}\cos{\left(\varphi-\phi\right)}}d\varphi}&=&\int_{0}^{2\pi}{\!\left[\alpha_{l,m}\sin{\varphi}e^{i\left(m\varphi-k\kappa_Rr\sin{\theta}\cos{\left(\varphi-\phi\right)}\right)}\!+\!im\beta_{l,m}e^{i\left[\left(m+1\right)\varphi-k\kappa_Rr\sin{\theta}\cos{\left(\varphi-\phi\right)}\right]}\right]d\varphi}\\
&=&\pi(-i)^{m}\left[\left(2m\beta_{l,m}-\alpha_{l,m}\right)J_{m+1}(k\kappa_Rr\sin{\theta})e^{i\phi}-\alpha_{l,m} J_{m-1}(k\kappa_Rr\sin{\theta})e^{-i\phi}\right]e^{im\phi},\\\\
\int_{0}^{2\pi}{\!\!\mathcal{Z}_{l,m}^{\rm (e)}e^{-ik\kappa_Rr\sin{\theta}\cos{\left(\varphi-\phi\right)}}d\varphi}&=&\int_{0}^{2\pi}{\!\gamma_{l,m}e^{i\left(m\varphi-k\kappa_Rr\sin{\theta}\cos{\left(\varphi-\phi\right)}\right)}d\varphi}=2\pi\left(-i\right)^m \gamma_{l,m}J_{m}(k\kappa_Rr\sin{\theta})e^{im\phi}.
\end{eqnarray*}
Even though these results appear to be somewhat complicated, we still have to carry out the integration along the radius and particularize to the case of the plane $z=0$. Thereupon, it turns out to be more convenient to express the latter integrals in cylindrical coordinates:
\begin{equation}
\tilde{{\bf \Pi}}^{\rm (e)}_{l,m}(\kappa_x,\kappa_y;z)=\frac{-k}{4\pi\omega}\frac{(-i)^mC_{l,m}}{\sqrt{l(l+1)}}\int_{0}^{+\infty}{\left\{\mathcal{\tilde{X}}_{l,m}^{\rm (e)}(R,z),\mathcal{\tilde{Y}}_{l,m}^{\rm (e)}(R,z),\mathcal{\tilde{Z}}_{l,m}^{\rm (e)}(R,z)\right\}dR},
\label{EqS84}
\end{equation}
where $R\equiv r\sin{\theta}$, $z\equiv r\cos{\theta}$, and
\begin{eqnarray}
\mathcal{\tilde{X}}_{l,m}^{\rm (e)}(R,z)&=&h_{l-1}(k\sqrt{R^2+z^2})R\left[\left(2im\beta_{l,m}-i\alpha_{l,m}\right) J_{m+1}(k\kappa_RR)e^{i\phi}+i\alpha_{l,m} J_{m-1}(k\kappa_RR)e^{-i\phi}\right]e^{im\phi},
\label{EqS85}\\\nonumber\\
\mathcal{\tilde{Y}}_{l,m}^{\rm (e)}(R,z)&=&h_{l-1}(k\sqrt{R^2+z^2})R\left[\left(2m\beta_{l,m}-\alpha_{l,m}\right)J_{m+1}(k\kappa_RR)e^{i\phi}-\alpha_{l,m} J_{m-1}(k\kappa_RR)e^{-i\phi}\right]e^{im\phi},
\label{EqS86}\\\nonumber\\
\mathcal{\tilde{Z}}_{l,m}^{\rm (e)}(R,z)&=&h_{l-1}(k\sqrt{R^2+z^2})R\left[2\gamma_{l,m}J_m(k\kappa_RR)\right]e^{im\phi}.
\label{EqS87}
\end{eqnarray}
\begin{table*}[t!]
\begin{tabular}{c | c c c}
\hline
\hline
 $\left\{l,m\right\}$ & $\mathcal{\tilde{X}}_{l,m}^{\rm (e)}(R,z)$ &  $\mathcal{\tilde{Y}}_{l,m}^{\rm (e)}(R,z)$ &   $\mathcal{\tilde{Z}}_{l,m}^{\rm (e)}(R,z)$\\

\hline
$\left\{1,0\right\}$ & $0$ &  $0$ &  $2Rh_0J_{0}$\\

$\left\{1,1\right\}$& $-2iRh_0J_0$ &  $2Rh_0J_0$ &  $0$\\
\hline & & & \\

$\left\{2,0\right\}$& $\displaystyle\left[
\frac{2iR^2h_1J_1}{\left(R^2+z^2\right)^{1/2}}\right]\cos{\phi}$ & $\displaystyle\left[ \frac{2iR^2h_1J_1}{\left(R^2+z^2\right)^{1/2}}\right]\sin{\phi}$ & $\displaystyle \frac{4Rzh_1J_0}{\left(R^2+z^2\right)^{1/2}}$\\

$\left\{2,1\right\}$& $\displaystyle \frac{-6iRzh_1J_0}{\left(R^2+z^2\right)^{1/2}}$ &  $\displaystyle \frac{6Rzh_1J_0}{\left(R^2+z^2\right)^{1/2}}$ &  $\displaystyle\left[ \frac{-6R^2h_1J_1}{\left(R^2+z^2\right)^{1/2}}\right]e^{i\phi}$\\

$\left\{2,2\right\}$& $\displaystyle \left[\frac{12iR^2h_1J_1}{\left(R^2+z^2\right)^{1/2}}\right]e^{i\phi}$&  $\displaystyle\left[ \frac{-12R^2h_1J_1}{\left(R^2+z^2\right)^{1/2}}\right]e^{i\phi}$ & $0$\\
& & & \\\hline& & & \\

$\left\{3,0\right\}$ & $\displaystyle\left[ \frac{6iR^2zh_2J_1}{R^2+z^2}\right]\cos{\phi}$ & $\displaystyle \left[\frac{6iR^2zh_2J_1}{R^2+z^2}\right]\sin{\phi}$ & $\displaystyle \frac{-3R
(R^2-2z^2)h_2J_0}{R^2+z^2}$\\
$\left\{3,1\right\}$& $\displaystyle \frac{3iRh_2\left[2(R^2-2z^2)J_0-R^2J_2e^{2i\phi}\right]}{R^2+z^2}$ & $\displaystyle \frac{-3Rh_2\left[2(R^2-2z^2)J_0+R^2J_2e^{2i\phi}\right]}{R^2+z^2}$ & $\displaystyle\left[ \frac{-24R^2zh_2J_1}{R^2+z^2}\right]e^{i\phi}$\\
$\left\{3,2\right\}$& $\displaystyle \left[\frac{60iR^2zh_2J_1}{R^2+z^2}\right]e^{i\phi}$ & $\displaystyle\left[ \frac{-60R^2zh_2J_1}{R^2+z^2}\right]e^{i\phi}$ & $\displaystyle \left[\frac{30R^3h_2J_2}{R^2+z^2}\right]e^{2i\phi}$\\
$\left\{3,3\right\}$& $\displaystyle \left[\frac{-90iR^3h_2J_2}{R^2+z^2}\right]e^{2i\phi}$ & $\displaystyle \left[\frac{90R^3h_2J_2}{R^2+z^2}\right]e^{2i\phi}$ & $0$\\
& & & \\\hline& & & \\

$\left\{4,0\right\}$& $\displaystyle \left[\frac{-3iR^2(R^2-4z^2)h_3J_1}{\left(R^2+z^2\right)^{3/2}}\right]\cos{\phi}$ & $\displaystyle \left[ \frac{-3iR^2(R^2-4z^2)h_3J_1}{\left(R^2+z^2\right)^{3/2}}\right]\sin{\phi}$ & $\displaystyle \frac{-4Rz
(3R^2-2z^2)h_3J_0}{\left(R^2+z^2\right)^{3/2}}$\\
$\left\{4,1\right\}$& $\displaystyle \frac{5iRzh_3\left[2(3R^2-2z^2)J_0-3R^2J_2e^{2i\phi}\right]}{\left(R^2+z^2\right)^{3/2}}$ &  $\displaystyle \frac{-5Rzh_3\left[2(3R^2-2z^2)J_0+3R^2J_2e^{2i\phi}\right]}{\left(R^2+z^2\right)^{3/2}}$ & $\displaystyle \left[ \frac{15R^2(R^2-4z^2)h_3J_1}{\left(R^2+z^2\right)^{3/2}}\right]e^{i\phi}$\\
$\left\{4,2\right\}$& $\displaystyle\left[ \frac{-15iR^2h_3\left[3(R^2-4z^2)J_1-R^2J_3e^{2i\phi}\right]}{\left(R^2+z^2\right)^{3/2}}\right]e^{i\phi}$ & $\displaystyle \left[ \frac{15R^2h_3\left[3(R^2-4z^2)J_1+R^2J_3e^{2i\phi}\right]}{\left(R^2+z^2\right)^{3/2}}\right]e^{i\phi}$ & $\displaystyle \left[\frac{180R^3zh_3J_2}{\left(R^2+z^2\right)^{3/2}}\right]e^{2i\phi}$\\
$\left\{4,3\right\}$& $\displaystyle \left[\frac{-630iR^3zh_3J_2}{\left(R^2+z^2\right)^{3/2}}\right]e^{2i\phi}$ & $\displaystyle \left[\frac{630R^3zh_3J_2}{\left(R^2+z^2\right)^{3/2}}\right]e^{2i\phi}$ & $\displaystyle \left[\frac{-210R^4h_3J_3}{\left(R^2+z^2\right)^{3/2}}\right]e^{3i\phi}$\\
$\left\{4,4\right\}$& $\displaystyle \left[\frac{840iR^4h_3J_3}{\left(R^2+z^2\right)^{3/2}}\right]e^{3i\phi}$ & $\displaystyle \left[ \frac{-840R^4h_3J_3}{\left(R^2+z^2\right)^{3/2}}\right]e^{3i\phi}$ & $0$\\
&&&\\\hline\hline
\end{tabular}
\caption{Spectral amplitudes of the Hertz potential associated to the first four electric-like multipole moments.}
\label{Table1}
\end{table*}

The explicit integrand functions given above are presented up to the fourth order in Table \ref{Table1}. From these results we can make an important observation: for each $l$-order, the different mathematical structures appearing in the $x$ and $y$ components (i.e., in the functions $\mathcal{\tilde{X}}_{l,m}^{\rm (e)}$ and $\mathcal{\tilde{Y}}_{l,m}^{\rm (e)}$), are actually linear combinations of the ones that appear in the $z$ component. Moreover, it can be seen that $\mathcal{\tilde{Z}}_{l,m}^{\rm (e)}(R,z)=0$ for all $l=m$. Therefore the problem is strongly reduced to the analysis of $l$ integrals associated to the $z$ component for each order $l$. Taking into account these considerations, in order to find a closed and analytical result, we shall first look into the possibility of integrating $\mathcal{\tilde{Z}}_{l,m}^{\rm (e)}(R,z)$ for these particular cases. This involves a judicious choice of integration by parts in those cases in which setting $z=0$ seems to cancel out the whole integral, greatly reminiscent of what happened in the previous case of the electric quadrupole. Thus, for the case $l=1$:
\begin{eqnarray}
\nonumber\int_{0}^{+\infty}{Rh_0(k\sqrt{R^2+z^2})P_0^0\left[\frac{z}{\sqrt{R^2+z^2}}\right]J_0(k\kappa_R R)dR}&=&\frac{-i}{k}\int_{0}^{+\infty}{\frac{Re^{ik\sqrt{R^2+z^2}}}{\sqrt{{R^2+z^2}}}J_0(k\kappa_R R)dR}\\
&\xrightarrow[z\to 0]{}&\frac{-i}{k}\int_0^{+\infty}{e^{ikR}J_0(k\kappa_R R)dR}=\frac{1}{k^2\kappa_z}.
\label{EqS88}
\end{eqnarray}

\noindent For $l=2$:
\begin{eqnarray}
\nonumber\int_{0}^{+\infty}{Rh_1(k\sqrt{R^2+z^2})P_1^1\left[\frac{z}{\sqrt{R^2+z^2}}\right]J_1(k\kappa_R R)dR}&=&\frac{1}{k}\int_{0}^{+\infty}{\left(1-\frac{1}{ik\sqrt{R^2+z^2}}\right)\frac{R^2e^{ik\sqrt{R^2+z^2}}}{R^2+z^2}J_1(k\kappa_R R)dR}\\
&\xrightarrow[z\to 0]{}&\frac{1}{k}\int_{0}^{+\infty}{\left(1-\frac{1}{ikR}\right)e^{ikR}J_1(k\kappa_R R)dR}=-\frac{\kappa_R}{k^2\kappa_z};
\label{EqS89}
\end{eqnarray}
\begin{eqnarray}
\nonumber\int_{0}^{+\infty}{Rh_1(k\sqrt{R^2+z^2})P_1^0\left[\frac{z}{\sqrt{R^2+z^2}}\right]J_0(k\kappa_R R)dR}&=&\int_{0}^{+\infty}{\left(\frac{1}{ik\sqrt{R^2+z^2}}-1\right)\frac{Rze^{ik\sqrt{R^2+z^2}}}{k(R^2+z^2)}J_0(k\kappa_RR)dR}\\
\nonumber&=&\left\{\begin{matrix*}[l]
u=J_0&\quad \Longrightarrow \quad & du=-k\kappa_RJ_1dR\\
dv=\displaystyle Rh_1P_1^1 dR&\quad \Longrightarrow \quad & v=\displaystyle-\frac{\sqrt{R^2+z^2}}{k}h_0P_1^0
\end{matrix*}\right\}\\
\nonumber&=&-\frac{i}{k^2}\frac{z}{|z|}e^{ik|z|}-\kappa_R\int_{0}^{+\infty}{\sqrt{R^2+z^2}h_0P_1^0J_1dR}\\
&\xrightarrow[z\to 0]{}&\mp \frac{i}{k^2}.
\label{EqS90}
\end{eqnarray}

\noindent For the case $l=3$ we have that:
\begin{eqnarray}
\nonumber\int_{0}^{+\infty}{Rh_2(k\sqrt{R^2+z^2})P_2^2\left[\frac{z}{\sqrt{R^2+z^2}}\right]J_2(k\kappa_ RR)dR}&=&\frac{i}{k}\int_{0}^{+\infty}{\!\!\left(3+\frac{9i}{k\sqrt{R^2+z^2}}-\frac{9}{k^2(R^2+z^2)}\right)\frac{R^3e^{ik\sqrt{R^2+z^2}}}{(R^2+z^2)^{3/2}}J_2dR}
\end{eqnarray}
\vspace{-0.4cm}
\begin{eqnarray}
\qquad\qquad\qquad\qquad\qquad\qquad\qquad\qquad\qquad\qquad&\xrightarrow[z\to 0]{}&\frac{i}{k}\int_{0}^{+\infty}{\!\!\left(3+\frac{9i}{kR}-\frac{9}{k^2R^2}\right)e^{ikR}J_2(k\kappa_RR)dR}=\frac{3\kappa_R^2}{k^2\kappa_z};
\label{EqS91}
\end{eqnarray}

\begin{eqnarray}
\nonumber\int_{0}^{+\infty}{Rh_2(k\sqrt{R^2+z^2})P_2^1\left[\frac{z}{\sqrt{R^2+z^2}}\right]J_1(k\kappa_R R)dR}&=&\frac{i}{k}\int_{0}^{+\infty}{\!\!\left(\frac{9}{k^2(R^2+z^2)}-\frac{9i}{k\sqrt{R^2+z^2}}-3\right)\frac{R^2ze^{ik\sqrt{R^2+z^2}}}{(R^2+z^2)^{3/2}}J_1dR}
\end{eqnarray}
\vspace{-0.4cm}
\begin{eqnarray}
\qquad\qquad\qquad\qquad\qquad\qquad\qquad\quad\nonumber&=&\left\{\begin{matrix*}[l]
u=RJ_1& \Longrightarrow  & du=k\kappa_RRJ_0dR\\
dv=\displaystyle h_2P_2^1 dR& \Longrightarrow  & v=\displaystyle \frac{3}{k}h_1P_1^0
\end{matrix*}\right\}\\
\nonumber&=&\left.\frac{3R}{k}h_1P_1^0J_1\right|_{R=0}^{R\to\infty}-3\kappa_R\int_{0}^{+\infty}{Rh_1P_1^0J_0dR}\\
&\xrightarrow[z\to 0]{}&\pm\frac{3i\kappa_R}{k^2};
\label{EqS92}
\end{eqnarray}

\begin{eqnarray}
\nonumber&&\int_{0}^{+\infty}{\!\!Rh_2(k\sqrt{R^2+z^2})P_2^0\left[\frac{z}{\sqrt{R^2+z^2}}\right]J_0(k\kappa_RR)dR} =\frac{i}{k}\int_{0}^{+\infty}{\!\!\left(\frac{3}{k^2\sqrt{R^2+z^2}}-\frac{3i}{k\sqrt{R^2+z^2}}-1\right)\frac{R(R^2-2z^2)e^{ik\sqrt{R^2+z^2}}}{2(R^2+z^2)^{3/2}}J_0dR}
\end{eqnarray}
\vspace{-0.4cm}
\begin{eqnarray}
\qquad\qquad\qquad\qquad\qquad\nonumber&=&\left\{\begin{matrix*}[l]
u=J_0& \Longrightarrow  & du=-k\kappa_RJ_1dR\\
dv=\displaystyle Rh_2P_2^0 dR& \Longrightarrow  & v=\displaystyle\left[2kz^2\sqrt{R^2+z^2}-R^2(3i+k\sqrt{R^2+z^2})\right]\frac{e^{ik\sqrt{R^2+z^2}}}{2k^3(R^2+z^2)^{3/2}}
\end{matrix*}\right\}\\
\nonumber&=&-\frac{e^{ik|z|}}{k^2}+\kappa_R\int_{0}^{+\infty}{\left[\frac{2kz^2\sqrt{R^2+z^2}-R^2(3i+k\sqrt{R^2+z^2})}{2k^2(R^2+z^2)^{3/2}}\right]e^{ik\sqrt{R^2+z^2}}J_1dR}\\
&\xrightarrow[z\to 0]{}&-\frac{1}{k^2}-\kappa_R\int_{0}^{+\infty}{\frac{(3i+k R)e^{ikR}}{2k^2R}J_1dR}=\frac{3\kappa_R^2-2}{2k^2\kappa_z}.
\label{EqS93}
\end{eqnarray}

\noindent And finally, for the case $l=4$:
\begin{eqnarray}
&&\int_{0}^{+\infty}{Rh_3(k\sqrt{R^2+z^2})P_3^3\left[\frac{z}{\sqrt{R^2+z^2}}\right]J_3(k\kappa_ RR)dR}\xrightarrow[z\to 0]{}\frac{-15\kappa_R^3}{k^2\kappa_z};
\label{EqS94}\\
&&\int_{0}^{+\infty}{Rh_3(k\sqrt{R^2+z^2})P_3^2\left[\frac{z}{\sqrt{R^2+z^2}}\right]J_2(k\kappa_ RR)dR}\xrightarrow[z\to 0]{}\mp\frac{15i\kappa_R^2}{k^2};
\label{EqS95}\\
&&\int_{0}^{+\infty}{Rh_3(k\sqrt{R^2+z^2})P_3^1\left[\frac{z}{\sqrt{R^2+z^2}}\right]J_1(k\kappa_ RR)dR}\xrightarrow[z\to 0]{}\frac{3\kappa_R(4-5\kappa_R^2)}{2k^2\kappa_z};
\label{EqS96}\\
&&\int_{0}^{+\infty}{Rh_3(k\sqrt{R^2+z^2})P_3^0\left[\frac{z}{\sqrt{R^2+z^2}}\right]J_0(k\kappa_ RR)dR}\xrightarrow[z\to 0]{}\mp\frac{i(5\kappa_R^2-2)}{2k^2}.
\label{EqS97}
\end{eqnarray}

After verifying that all the integrands showed in Table \ref{Table1} are actually integrable functions, we are in the position to derive a general formula enabling this integration for all the possible cases of $l$ and $m$. To this aim, we first observe that the above integrals involving the radial variable all have the same form:
\begin{equation}
\mathcal{K}_{l,m}^{l',m'}(k,\kappa_R;z)=\int_{0}^{+\infty}{Rh_{l-1}{\left(k\sqrt{R^2+z^2}\right)}P_{l'}^{m'}{\left[\frac{z}{\sqrt{R^2+z^2}}\right]}J_m{\left(k\kappa_R R\right)}dR}.
\label{EqS98}
\end{equation}
It is important to stress that the latter integral involves the product of both spherical and cylindrical Bessel-like functions. This fact hinders the direct utilization of usual integral properties involving Bessel functions. We may yet write the spherical Bessel functions of first and second kinds in terms of the (cylindrical) Bessel functions of first kind and half-integer order \cite{Watsonsm}:
\begin{equation}
j_l(kr)=\sqrt{\frac{\pi}{2 kr}}J_{l+\frac{1}{2}}(kr),\qquad\qquad y_l(kr)=\sqrt{\frac{\pi}{2 kr}}Y_{l+\frac{1}{2}}(kr)=(-1)^{l+1}\sqrt{\frac{\pi}{2kr}}J_{-l-\frac{1}{2}}(kr).
\label{EqS99}
\end{equation}
So, the spherical Hankel functions can be expressed as 
\begin{equation}
h_{l}(kr)=j_l(kr)+iy_l(kr)=\sqrt{\frac{\pi}{2kr}}\left[J_{l+\frac{1}{2}}(kr)+i(-1)^{l+1}J_{-l-\frac{1}{2}}(kr)\right],
\label{EqS100}
\end{equation}
and the integral given in Eq. \eqref{EqS98} is then recast as
\begin{equation}
\mathcal{K}_{l,m}^{l',m'}(k,\kappa_R;z)=\sqrt{\frac{\pi}{2k}}\int_{0}^{+\infty}{\frac{R}{\left(R^2+z^2\right)^{1/4}}\left[J_{l-\frac{1}{2}}+i(-1)^{l}J_{-l+\frac{1}{2}}\right]P_{l'}^{m'}J_{m}dR},
\label{EqS101}
\end{equation}
where we have omitted the argument in both the Bessel and associated Legendre functions. Furthermore, since we are ultimately considering the plane $z=0$, according to the results given in Eq. \eqref{EqS79} we can take the Legendre functions outside the integral, and then, it turns out to be convenient to define $\mathcal{K}_{l,m}^{l',m'}(k,\kappa_R;z)\equiv P_{l'}^{m'}(z)\mathcal{I}_{l,m}(k,\kappa_R;z)$. Under this condition, it can be explicitly demonstrated (by means of a symbolic calculation software, e.g., Wolfram \textit{Mathematica}) that Eq. \eqref{EqS101} yields a closed result:
\begin{eqnarray}
\!\!\!\!\!\!\!\!\!\!\!\!\!\!\!\!\mathcal{I}_{l,m}^{\rm (odd)}(k,\kappa_R;0)=\sqrt{\frac{\pi}{2k}}\int_{0}^{+\infty}{\!\!\!\!\sqrt{R}J_{l-\frac{1}{2}}J_{m}dR}&=&\frac{\sqrt{\pi}\kappa_R^m}{k^{2}}\frac{\Gamma\left[\frac{1+l+m}{2}\right]}{\Gamma\left[\frac{l-m}{2}\right]} \frac{{}_2F_1\left[\frac{1}{2}\left(2-l+m\right),\frac{1}{2}\left(1+l+m\right),1+m;\kappa_R^2\right]}{\Gamma\left[1+m\right]},
\label{EqS102}\\
\!\!\!\!\!\!\!\!\!\!\!\!\!\!\!\!\mathcal{I}_{l,m}^{\rm (even)}(k,\kappa_R;0)=\sqrt{\frac{\pi}{2k}}\int_{0}^{+\infty}{\!\!\!\!\sqrt{R}J_{\frac{1}{2}-l}J_{m}dR}&=&\frac{\sqrt{\pi}\kappa_R^m}{k^{2}}\frac{\Gamma\left[\frac{2-l+m}{2}\right]}{\Gamma\left[\frac{1-l-m}{2}\right]} \frac{{}_2F_1\left[\frac{1}{2}\left(2-l+m\right),\frac{1}{2}\left(1+l+m\right),1+m;\kappa_R^2\right]}{\Gamma\left[1+m\right]},
\label{EqS103}
\end{eqnarray}
where ${}_2F_1(a,b,c;z)$ is the Gaussian hypergeometric function defined as \cite{Abramowitzsm}
\begin{equation}
{}_2F_1(a,b,c;z)\equiv\frac{\Gamma{[c]}}{\Gamma{[a]}\Gamma{[b]}}\sum_{t=0}^{\infty}{\frac{\Gamma{[a+t]}\Gamma{[b+t]}}{\Gamma{[c+t]}}\frac{z^t}{t!}}.
\label{EqS104}
\end{equation}
Hence, by summing up Eqs. \eqref{EqS102} and \eqref{EqS103}, it follows that
\begin{tcolorbox}[colback=blue!7,colframe=white!80!black]
\vspace{-0.25cm}
\begin{equation}
\!\!\!\!\!\!\!\!\mathcal{I}_{l,m}(k,\kappa_R;0)=\frac{\sqrt{\pi}\kappa_R^m}{k^2}\frac{{}_2F_1\left[\frac{1}{2}\left(2-l+m\right),\frac{1}{2}\left(1+l+m\right),1+m,\kappa_R^2\right]}{\Gamma\left[1+m\right]} \left[\frac{\Gamma\left[\frac{1+l+m}{2}\right]}{\Gamma\left[\frac{l-m}{2}\right]}+i\left(-1\right)^l\frac{\Gamma\left[\frac{2-l+m}{2}\right]}{\Gamma\left[\frac{1-l-m}{2}\right]}\right].
\label{EqS105}
\end{equation}
\end{tcolorbox}
\noindent Looking at this result carefully one realizes that there are two situations that should be distinguished when $z\to0$. On the one hand, if $l+m$ is odd, $\mathcal{I}_{l,m}^{\rm (odd)}$ is a well behaved function of $k$ and $\kappa_R$ (indeed this happens regardless of the parity of $l+m$) and $\mathcal{I}_{l,m}^{\rm (even)}=0$. However, if $l+m$ is even the function $\mathcal{I}_{l,m}^{\rm (even)}$ strikingly blows up. As pointed out by Mandel regarding the simplest case of the spherical wave $e^{ikr}/r$ (when addressing the classical Weyl's identity) \cite{Mandelsm}, such divergent behavior is a direct consequence of the presence of a singularity at the origin. Still, as shown above, we can perform the integration by hand, thereby indicating that these singularities are actually removable. In fact, our calculations up to order four show that in the limit $z\to0$, the infinities of $\mathcal{I}_{l,m}$ are compensated by the zeros of $P_{l'}^{m'}$ \cite{footnote2sm} in such a way that the function $\mathcal{K}_{l,m}^{l',m'}(k,\kappa_R;z)$ can always be evaluated for whatever order $(l,m)$, thus leading to a well defined function of $k$ and $\kappa_R$. Hence, the integral given in Eq. \eqref{EqS98} ought to be redefined as a limit \cite{footnote3sm}:
\begin{tcolorbox}[colback=red!10,colframe=white!80!black]
\vspace{-0.25cm}
\begin{equation}
\tilde{\mathcal{K}}_{l,m}^{l',m'}(k,\kappa_R)\equiv\lim\limits_{z\to0}{\mathcal{K}_{l,m}^{l',m'}(k,\kappa_R;z)}=\lim\limits_{z\to0}{P_{l'}^{m'}(z)\mathcal{I}_{l,m}(k,\kappa_R;z)}.
\label{EqS106}
\end{equation}
\end{tcolorbox}
\clearpage 
Putting it all together, the spectral amplitude of the Hertz vector potential associated to electric-like sources of order $(l,m)$ on the plane $z=0$ finally reads as
\begin{tcolorbox}[colback=blue!7,colframe=white!80!black]
\vspace{-0.25cm}
\begin{equation}
\!\!\!\!\tilde{\bf \Pi}^{\rm (e)}_{l,m}(\kappa_x,\kappa_y;z\to 0)\!=\!\frac{-k}{4\pi\omega}\frac{(-i)^mC_{l,m}}{\sqrt{l(l+1)}}\!\left[\tilde{\Pi}_{l,m}^{\rm (e),x}(\kappa_x,\kappa_y;z\to0),\tilde{\Pi}_{l,m}^{\rm (e),y}(\kappa_x,\kappa_y;z\to0),\tilde{\Pi}_{l,m}^{\rm (e),z}(\kappa_x,\kappa_y;z\to0)\right]\!\!,
\label{EqS107}
\end{equation}
where
\begin{eqnarray*}
\tilde{\Pi}_{l,m}^{\rm(e),x}(\kappa_x,\kappa_y;z\to0)&=&i\left[(m-l)\tilde{\mathcal{K}}_{l,m+1}^{l,m}(k,\kappa_R)e^{i\phi}+(m+l)\tilde{\mathcal{K}}_{l,m-1}^{l,m}(k,\kappa_R)e^{-i\phi}\right]e^{im\phi},\\
\tilde{\Pi}_{l,m}^{\rm(e),y}(\kappa_x,\kappa_y;z\to0)&=&\left[(m-l)\tilde{\mathcal{K}}_{l,m+1}^{l,m}(k,\kappa_R)e^{i\phi}-(m+l)\tilde{\mathcal{K}}_{l,m-1}^{l,m}(k,\kappa_R)e^{-i\phi}\right]e^{im\phi},\\
\tilde{\Pi}_{l,m}^{\rm(e),z}(\kappa_x,\kappa_y;z\to0)&=&2(l+m)\tilde{\mathcal{K}}_{l,m}^{l-1,m}(k,\kappa_R)e^{im\phi}.
\end{eqnarray*}
\end{tcolorbox}
\noindent And those of magnetic sources can be readily obtained from these, as detailed in Eq. \eqref{EqS76}. Then, using this general expression, we can analytically obtain the angular spectra for the Hertz potential of arbitrary multipoles. For the first four orders in $l$, these are given by:
\begin{tcolorbox}[colback=blue!7,colframe=white!80!black]
\vspace{-0.25cm}
\begin{eqnarray*}
\tilde{{\bf \Pi}}^{\rm (e)}_{1,0}(\kappa_x,\kappa_y;z\to0)&=& \frac{-C_{1,0}}{4\sqrt{2}\pi}\frac{2}{k\omega}\frac{1}{\kappa_z}\left\{0,0,1\right\},\\
\tilde{{\bf \Pi}}^{\rm (e)}_{1,1}(\kappa_x,\kappa_y;z\to0)&=& \frac{iC_{1,1}}{4\sqrt{2}\pi}\frac{4}{k\omega}\frac{1}{\kappa_z}\frac{1}{2}\left\{-i,1,0\right\};\\\\\\
\tilde{{\bf \Pi}}^{\rm (e)}_{2,0}(\kappa_x,\kappa_y;z\to0)&=& \frac{-C_{2,0}}{4\sqrt{6}\pi}\frac{4}{k\omega}\frac{1}{\kappa_z}\frac{1}{2}\left\{i\kappa_x,i\kappa_y,\mp 2i\kappa_z\right\},\\
\tilde{{\bf \Pi}}^{\rm (e)}_{2,1}(\kappa_x,\kappa_y;z\to0)&=& \frac{iC_{2,1}}{4\sqrt{6}\pi}\frac{6}{k\omega}\frac{1}{\kappa_z}\left\{\mp \kappa_z,\mp i\kappa_z,-(\kappa_x+i\kappa_y)\right\},\\
\tilde{{\bf \Pi}}^{\rm (e)}_{2,2}(\kappa_x,\kappa_y;z\to0)&=& \frac{C_{2,2}}{4\sqrt{6}\pi}\frac{8}{k\omega}\frac{1}{\kappa_z}\frac{3}{2}\left\{i(\kappa_x+i\kappa_y),-(\kappa_x+i\kappa_y),0\right\};\\\\\\
\tilde{{\bf \Pi}}^{\rm (e)}_{3,0}(\kappa_x,\kappa_y;z\to0)&=& \frac{-C_{3,0}}{4\sqrt{12}\pi}\frac{6}{k\omega}\frac{1}{\kappa_z}\frac{1}{2}\left\{\pm 2\kappa_x\kappa_z,\pm2 \kappa_y\kappa_z,3(\kappa_x^2+\kappa_y^2)-2\right\},\\
\tilde{{\bf \Pi}}^{\rm (e)}_{3,1}(\kappa_x,\kappa_y;z\to0)&=& \frac{iC_{3,1}}{4\sqrt{12}\pi}\frac{8}{k\omega}\frac{1}{\kappa_z}\frac{3}{8}\left\{-i(7\kappa_x^2+5\kappa_y^2+2i\kappa_x\kappa_y-4),5\kappa_x^2+7\kappa_y^2-2i\kappa_x\kappa_y-4,\pm 8i \kappa_z(\kappa_x+i\kappa_y)\right\},\\
\tilde{{\bf \Pi}}^{\rm (e)}_{3,2}(\kappa_x,\kappa_y;z\to0)&=& \frac{C_{3,2}}{4\sqrt{12}\pi}\frac{10}{k\omega}\frac{1}{\kappa_z}3\left\{\pm 2(\kappa_x+i\kappa_y)\kappa_z,\pm 2i(\kappa_x+i\kappa_y)\kappa_z,(\kappa_x+i\kappa_y)^2\right\},\\
\tilde{{\bf \Pi}}^{\rm (e)}_{3,3}(\kappa_x,\kappa_y;z\to0)&=& \frac{-iC_{3,3}}{4\sqrt{12}\pi}\frac{12}{k\omega}\frac{1}{\kappa_z}\frac{15}{2}\left\{-i(\kappa_x+i\kappa_y)^2,(\kappa_x+i\kappa_y)^2,0\right\};\\\\\\
\tilde{{\bf \Pi}}^{\rm (e)}_{4,0}(\kappa_x,\kappa_y;z\to0)&=& \frac{-C_{4,0}}{4\sqrt{20}\pi}\frac{8}{k\omega}\frac{1}{\kappa_z}\frac{1}{8}\left\{3i(5\kappa_R^2-4)\kappa_R\cos{\phi},3i(5\kappa_R^2-4)\kappa_R\sin{\phi},\pm 4i\kappa_z(2-5\kappa_R^2)\right\},\\
\tilde{{\bf \Pi}}^{\rm (e)}_{4,1}(\kappa_x,\kappa_y;z\to0)&=& \frac{iC_{4,1}}{4\sqrt{20}\pi}\frac{10}{k\omega}\frac{1}{\kappa_z}\frac{1}{2}\left\{\pm\kappa_z\left[(10+3e^{2i\phi})\kappa_R^2-4\right],\pm i\kappa_z\left[(10-3e^{2i\phi})\kappa_R^2-4\right],3(4-5\kappa_R^2)\kappa_Re^{i\phi}\right\},\\
\tilde{{\bf \Pi}}^{\rm (e)}_{4,2}(\kappa_x,\kappa_y;z\to0)&=& \frac{C_{4,2}}{4\sqrt{20}\pi}\frac{12}{k\omega}\frac{1}{\kappa_z}\frac{5}{4}\left\{i\left[(15+e^{2i\phi})\kappa_R^2-12\right]\kappa_Re^{i\phi},\left[(e^{2i\phi}-15)\kappa_R^2+12\right]\kappa_Re^{i\phi},\mp12i \kappa_z\kappa_R^2e^{2i\phi}\right\},\\
\tilde{{\bf \Pi}}^{\rm (e)}_{4,3}(\kappa_x,\kappa_y;z\to0)&=& \frac{-iC_{4,3}}{4\sqrt{20}\pi}\frac{14}{k\omega}\frac{1}{\kappa_z}15\left\{\mp 3\kappa_z\kappa_R^2e^{2i\phi},\mp 3i\kappa_z\kappa_R^2e^{2i\phi},-\kappa_R^3e^{3i\phi}\right\},\\
\tilde{{\bf \Pi}}^{\rm (e)}_{4,4}(\kappa_x,\kappa_y;z\to0)&=& \frac{-C_{4,4}}{4\sqrt{20}\pi}\frac{16}{k\omega}\frac{1}{\kappa_z}\frac{105}{2}\left\{i\kappa_R^3e^{3i\phi},-\kappa_R^3e^{3i\phi},0\right\}.
\end{eqnarray*}
\end{tcolorbox}

As a final remark, it should be noted that a similar derivation of the angular spectra of EM multipole fields of arbitrary order has already been addressed with other approaches \cite{Devaney1974sm,Bobbert1986sm,Borghi2004sm}. Yet, even though the final results are explicitly expressed by means of an analytical and closed form (see, e.g., those shown in Ref. \cite{Borghi2004sm}), both the formalism and the notation used makes them hard to interpret and handle, thus being not very practical.

\section{VI. Verification and relationship between the results obtained via the Hertz's and standard vector potentials for both the electric dipole and quadrupole}

Below, in order to verify our results, we will compare the above general expression of the spectral amplitude [Eq. \eqref{EqS107}] for the particular cases of the electric dipole ($l=1$) and quadrupole ($l=2$), with those explicitly obtained in Secs.$\,\,$II and III [Eqs. \eqref{EqS21} and \eqref{EqS37}], respectively. 

In the first place, focusing on the electric dipole, we have to consider $l=1$ and $m=0,\pm1$, which correspond to the different characteristic polarizations of the electric dipole moment ${\bf p}$. To determine their explicit relationship we should look into the EM fields. Specifically, just by looking at the magnetic field stemmed from the standard vector potential of the electric dipole given in Eq. \eqref{EqS17}, we have that
\begin{equation}
\nonumber{\bf H}^{\rm ED}=\frac{\omega k^2}{4\pi}h_1(kr)\left\{\left[p_y\cos{\varphi}-p_x\sin{\varphi}\right]{\bf e}_\theta+\left[p_z\sin{\theta}-\cos{\theta}\left(p_x\cos{\varphi}+p_y\sin{\varphi}\right)\right]{\bf e}_\varphi\right\},
\end{equation}
where, for comparison purposes, we have expressed the dipole moment in the spherical basis. This magnetic field should be related to the one derived from the multipole expansion for $l=1$:
\begin{equation}
{\bf H}_{1,m}^{\rm (e)}=-\frac{irh_1(kr)}{\sqrt{2}}\left[{\bf e}_r\times\nabla Y_{1,m}(\Omega)\right] \Rightarrow \left\{\begin{matrix*}[l]
m=0 & \rightarrow & {\bf H}_{1,0}^{\rm (e)}\;\:=\displaystyle\frac{ih_1(kr)}{\sqrt{2}}\sqrt{\frac{3}{4\pi}}\sin{\theta}{\bf e}_\varphi,\\\\
m=\pm 1& \rightarrow & {\bf H}_{1,\pm 1}^{\rm (e)}=\displaystyle\frac{ih_1(kr)}{\sqrt{2}}\sqrt{\frac{3}{8\pi}}\left(-i{\bf e}_\theta\pm\cos{\theta}{\bf e}_\varphi\right)e^{\pm i\varphi}.
\end{matrix*}\right.
\label{EqS108}
\end{equation}
Then, leaving aside the prefactors, it is easy to see that, for $m=0$, the functional form of ${\bf H}^{\rm ED}$ coincides with that of ${\bf H}^{\rm (e)}_{1,m}$  as long as $p_x=p_y=0$. Likewise, for $m=\pm 1$, the matching is satisfied for $p_x=\mp ip_y$ and $p_z=0$. Taking into account these relations, one can directly observe that:
\begin{eqnarray}
{\bf H}^{\rm ED}=\frac{\omega k^2}{4\pi}\frac{(-1)^m\sqrt{2}}{iC_{1,m}}{\bf H}_{1,m}^{\rm (e)}\Rightarrow \left\{\begin{matrix*}[l]
{\bf H}^{\rm ED}&=&-\displaystyle\frac{\omega k^2}{4\pi}h_1(kr)\left[{\bf e}_r\times{\bf p}_0\right]=\frac{\omega k^2}{4\pi}\frac{\sqrt{2}}{iC_{1,0}}{\bf H}_{1,0}^{\rm (e)},\\\\
{\bf H}^{\rm ED}&=&-\displaystyle\frac{\omega k^2}{4\pi}h_1(kr)\left[{\bf e}_r\times{\bf p}_\pm\right]=\frac{\omega k^2}{4\pi}\frac{(-1)\sqrt{2}}{iC_{1,\pm1}}{\bf H}_{1,\pm1}^{\rm (e)},
\end{matrix*}\right.
\label{EqS109}
\end{eqnarray}
where $C_{1,0}=\sqrt{3/4\pi}$, $C_{1,+1}=\sqrt{3/8\pi}$, and $C_{1,-1}=(-1)C_{1,+1}$. Therefore it can be shown that for $l=1$, the following relationship holds:
\begin{equation}
\begin{matrix*}[l]
\left\{l,m\right\}=\left\{1,0\right\}&\quad\Longleftrightarrow\quad& {\bf p}_0=\left\{0,0,1\right\},\\\\
\left\{l,m\right\}=\left\{1,\pm1 \right\}&\quad\Longleftrightarrow\quad& {\bf p}_\pm=\left\{1,\pm i,0\right\}.
\end{matrix*}
\label{EqS110}
\end{equation}
With this in mind, we can directly compare the result shown in Eq. \eqref{EqS21} with that given in Eq. \eqref{EqS107} for $l=1$:
\begin{equation*}
\begin{aligned}
&\tilde{{\bf A}}^{\rm ED}_{0}=\frac{\mu ck^2}{8\pi^2n}\frac{1}{\kappa_z}{\bf p}_0 &&  \Leftrightarrow &&\tilde{{\bf \Pi}}^{\rm (e)}_{1,0}=\frac{-C_{1,0}}{2\sqrt{2}\pi}\frac{1}{k\omega}\frac{1}{\kappa_z}{\bf p}_0  && \Rightarrow &&\tilde{{\bf A}}^{\rm ED}_0=\frac{-\sqrt{2}}{4\pi C_{1,0}}\frac{\mu c k^3\omega}{n}\tilde{{\bf \Pi}}^{\rm (e)}_{1,0},\\\\
&\tilde{{\bf A}}^{\rm ED}_{\pm1}=\frac{\mu ck^2}{8\pi^2n}\frac{1}{\kappa_z}{\bf p}_{\pm} && \Leftrightarrow &&\tilde{{\bf \Pi}}^{\rm (e)}_{1,\pm1}=\frac{C_{1,\pm1}}{2\sqrt{2}\pi}\frac{1}{k\omega}\frac{1}{\kappa_z}{\bf p}_{\pm } && \Rightarrow &&\tilde{{\bf A}}^{\rm ED}_{\pm1}=\frac{\sqrt{2}}{4\pi C_{1,\pm1}}\frac{\mu c k^3\omega}{n}\tilde{{\bf \Pi}}^{\rm (e)}_{1,\pm1},
\end{aligned}
\end{equation*}
where it should be noted that $\tilde{{\bf A}}_{-1}^{\rm ED}=[\tilde{{\bf A}}_{+1}^{\rm ED}]^*$ and $\tilde{{\bf \Pi}}_{1,-1}^{(e)}=[-\tilde{{\bf \Pi}}_{1,+1}^{(e)}]^*$. Hence, for this particular case, apart from the prefactors that account for the dimensional constants, both approaches are in perfect agreement, as expected.

Following a similar procedure for the electric quadrupole, we need first to identify the explicit relationship between the components of the quadrupole moment tensor, $\dvec{\mathcal{Q}}$, and the different pairs $\left\{l,m\right\}$, with $l=2$ and $m=0,\pm 1,\pm2$. For this, we will write the magnetic field associated to the electric quadrupole from the corresponding standard vector potential given in Eq. \eqref{EqS25}:
\begin{eqnarray*}
{\bf H}^{\rm EQ}&=&\frac{\omega k^3}{48\pi}h_2(kr)\left[2\cos{\theta}(\mathcal{Q}_{yz}\cos{\varphi}-\mathcal{Q}_{xz}\sin{\varphi})+\sin{\theta}(2\mathcal{Q}_{xy}\cos{(2\varphi)}+(\mathcal{Q}_{yy}-\mathcal{Q}_{xx})\sin{(2\varphi)})\right]{\bf e}_\theta\\
&&\!\!\!\!\!\!\!\!\!\!\!\!\!\!\!\!\!\!\!\!\!-\frac{\omega k^3}{96\pi}h_2(kr)\left\{4\cos{(2\theta)}(\mathcal{Q}_{xz}\cos{\varphi}+\mathcal{Q}_{yz}\sin{\varphi})+\sin{(2\theta)}\left[3(\mathcal{Q}_{xx}+\mathcal{Q}_{yy})+(\mathcal{Q}_{xx}-\mathcal{Q}_{yy})\cos{(2\varphi)}+2\mathcal{Q}_{xy}\sin{(2\varphi)}\right]\right\}{\bf e}_\varphi.
\end{eqnarray*}
Analogously to the dipole case, this expression of the magnetic field has to be related to that obtained from the multipole expansion for $l=2$:
\vspace{-0.2cm}
\begin{equation}
\!\!\!{\bf H}_{2,m}^{\rm (e)}=-\frac{irh_2(kr)}{\sqrt{6}}\left[{\bf e}_r\times\nabla Y_{2,m}(\Omega)\right]\Rightarrow \left\{\begin{matrix*}[l]
m=0 & \rightarrow & {\bf H}_{2,0}^{\rm (e)}\:=\:\displaystyle\frac{ih_2(kr)}{\sqrt{6}}\frac{3}{4}\sqrt{\frac{5}{\pi}}\sin{(2\theta)}{\bf e}_\varphi,\\\\
m=\pm 1& \rightarrow & {\bf H}_{2,\pm 1}^{\rm (e)}\!=\displaystyle\frac{ih_2(kr)}{\sqrt{6}}\sqrt{\frac{15}{8\pi}}\left[-i\cos{\theta}{\bf e}_\theta\pm \cos{(2\theta)}{\bf e}_\varphi\right]e^{\pm i\varphi},\\\\
m=\pm 2& \rightarrow & {\bf H}_{2,\pm 2}^{\rm (e)}=\displaystyle\frac{ih_2(kr)}{\sqrt{6}}\sqrt{\frac{15}{32\pi}}\left[\pm2i\sin{\theta}{\bf e}_\theta-\sin{(2\theta)}{\bf e}_\varphi\right]e^{\pm 2i\varphi}.
\end{matrix*}\right.
\label{EqS111}
\end{equation}
Thus, by comparing the functional form of ${\bf H}^{\rm EQ}$ with that of ${\bf H}_{2,m}^{\rm (e)}$, it is easy to see that, for $m=0$ we have to require that $\mathcal{Q}_{xx}=\mathcal{Q}_{yy}$, and $\mathcal{Q}_{yz}=\mathcal{Q}_{xz}=\mathcal{Q}_{xy}=0$. Likewise, for $m=\pm1$, $\mathcal{Q}_{xx}=\mathcal{Q}_{yy}=\mathcal{Q}_{xy}=0$, and $\mathcal{Q}_{yz}=\pm i\mathcal{Q}_{xz}$. And finally, for $m=\pm 2$ we have that $\mathcal{Q}_{xy}=\pm i\mathcal{Q}_{xx}=\mp i\mathcal{Q}_{yy}$, and $\mathcal{Q}_{xz}=\mathcal{Q}_{yz}=0$. With these conditions, it follows that
\begin{eqnarray}
{\bf H}^{\rm EQ}=\frac{\omega k^2}{24\pi}\frac{(-1)^m\sqrt{6}}{iC_{2,m}}{\bf H}_{2,m}^{\rm (e)}\Rightarrow \left\{\begin{matrix*}[l]
{\bf H}^{\rm EQ}&=&-\displaystyle\frac{\omega k^3}{24\pi}h_2(kr)\left[{\bf e}_r\times\left(\dvec{\mathcal{Q}}_0\cdot{\bf e}_r\right)\right]=\frac{\omega k^2}{24\pi}\frac{\sqrt{6}}{iC_{2,0}}{\bf H}_{2,0}^{\rm (e)},\\\\
{\bf H}^{\rm EQ}&=&-\displaystyle\frac{\omega k^3}{24\pi}h_2(kr)\left[{\bf e}_r\times\left(\dvec{\mathcal{Q}}_{\pm1}\cdot{\bf e}_r\right)\right]=\frac{\omega k^2}{24\pi}\frac{(-1)\sqrt{6}}{iC_{2,\pm1}}{\bf H}_{2,\pm 1}^{\rm (e)},\\\\
{\bf H}^{\rm EQ}&=&-\displaystyle\frac{\omega k^3}{24\pi}h_2(kr)\left[{\bf e}_r\times\left(\dvec{\mathcal{Q}}_{\pm2}\cdot{\bf e}_r\right)\right]=\frac{\omega k^2}{24\pi}\frac{\sqrt{6}}{iC_{2,\pm2}}{\bf H}_{2,\pm 2}^{\rm (e)},
\end{matrix*}\right.
\label{EqS112}
\end{eqnarray}
where $C_{2,0}=\sqrt{5/4\pi}$, $C_{2,1}=\sqrt{5/24\pi}$, $C_{2,2}=\sqrt{5/96\pi}$, and $C_{2,-m}=(-1)^mC_{2,m}$. Therefore, for $l=2$, we get the following correspondence between the $m$ multipolar order and the matrices associated to the quadrupole moment tensor (expressed in Cartesian basis):
\vspace{-0.2cm}
\begin{equation}
\begin{matrix*}[l]
\left\{l,m\right\}=\left\{2,0\right\}&\qquad\Longleftrightarrow\qquad&\dvec{\mathcal{Q}}_0=\displaystyle\frac{1}{k}
\begin{pmatrix}
-1	& 	0	&	0\\
0	& 	-1	&	0\\
0	& 	0	&	2
\end{pmatrix},\\\\
\left\{l,m\right\}=\left\{2,\pm1\right\}&\qquad\Longleftrightarrow\qquad&\dvec{\mathcal{Q}}_{\pm 1}=\displaystyle\frac{3}{k}
\begin{pmatrix}
0	& 	0	&	1\\
0	& 	0	&	\pm i\\
1	& 	\pm i	&	0
\end{pmatrix},\\\\
\left\{l,m\right\}=\left\{2,\pm 2\right\}&\qquad\Longleftrightarrow\qquad&\dvec{\mathcal{Q}}_{\pm 2}=\displaystyle\frac{6}{k}
\begin{pmatrix}
\pm 1	& 	i	&	0\\
i	& 	\mp 1	&	0\\
0	& 	0	&	0
\end{pmatrix}.
\end{matrix*}
\label{EqS113}
\end{equation}
Taking into account these latter relations we can now compare the particular result of Eq. \eqref{EqS37} with the general one given in Eq. \eqref{EqS107} for $l=2$:
\begin{eqnarray*}
\begin{aligned}
&\!\!\!\!\!\!\!\!\!\!\!\!\!\!\!\!\!\!\!\!\!\!\!\!\tilde{{\bf A}}^{\rm EQ}_{0}=\frac{\mu ck^2}{8\pi^2n}\frac{(-i)}{6\kappa_z}\left[\dvec{\mathcal{Q}}_{0}\cdot{\bf k}^\pm\right] && \!\!\!\! \Leftrightarrow &&\!\!\tilde{{\bf \Pi}}^{\rm (e)}_{2,0}=\frac{iC_{2,0}}{2\sqrt{6}\pi}\frac{1}{k\omega}\frac{1}{\kappa_z}\left[\dvec{\mathcal{Q}}_0\cdot{\bf k}^\pm\right]  &&\!\!\!\!\!\! \Rightarrow &&\!\!\tilde{{\bf A}}^{\rm EQ}_0=\frac{-\sqrt{6}}{24\pi C_{2,0}}\frac{\mu c k^3\omega}{n}\tilde{{\bf \Pi}}^{\rm (e)}_{2,0},\\\\
\end{aligned}
\end{eqnarray*}
\begin{eqnarray*}
\begin{aligned}
&\!\!\!\!\!\!\!\!\!\!\!\!\!\!\!\!\!\!\!\!\!\!\!\!\tilde{{\bf A}}^{\rm EQ}_{\pm1}=\frac{\mu ck^2}{8\pi^2n}\frac{(-i)}{6\kappa_z}\left[\dvec{\mathcal{Q}}_{\pm 1}\cdot{\bf k}^\pm\right] &&\!\!\!\!  \Leftrightarrow &&\!\!\tilde{{\bf \Pi}}^{\rm (e)}_{2,\pm1}=\frac{\mp iC_{2,\pm 1}}{2\sqrt{6}\pi}\frac{1}{k\omega}\frac{1}{\kappa_z}\left[\dvec{\mathcal{Q}}_{\pm 1}\cdot{\bf k}^\pm\right]  &&\!\!\!\!\!\! \Rightarrow  &&\!\!\tilde{{\bf A}}^{\rm EQ}_{\pm 1}=\frac{\pm\sqrt{6}}{24\pi C_{2,\pm1}}\frac{\mu c k^3\omega}{n}\tilde{{\bf \Pi}}^{\rm (e)}_{2,\pm 1},\\\\
&\!\!\!\!\!\!\!\!\!\!\!\!\!\!\!\!\!\!\!\!\!\!\!\!\tilde{{\bf A}}^{\rm EQ}_{\pm2}=\frac{\mu ck^2}{8\pi^2n}\frac{(-i)}{6\kappa_z}\left[\dvec{\mathcal{Q}}_{\pm 2}\cdot{\bf k}^\pm\right] &&\!\!\!\!  \Leftrightarrow &&\!\!\tilde{{\bf \Pi}}^{\rm (e)}_{2,\pm 2}=\frac{iC_{2,\pm 2}}{2\sqrt{6}\pi}\frac{1}{k\omega}\frac{1}{\kappa_z}\left[\dvec{\mathcal{Q}}_{\pm 2}\cdot{\bf k}^\pm\right]  &&\!\!\!\!\!\! \Rightarrow  &&\!\!\tilde{{\bf A}}^{\rm EQ}_{\pm2}=\frac{-\sqrt{6}}{24\pi C_{2,\pm2}}\frac{\mu c k^3\omega}{n}\tilde{{\bf \Pi}}^{\rm (e)}_{2,\pm 2},
\end{aligned}
\end{eqnarray*}
with $\tilde{{\bf A}}_{-1}^{\rm EQ}=[-\tilde{{\bf A}}_{+1}^{\rm EQ}]^*$, $\tilde{{\bf A}}_{-2}^{\rm EQ}=[\tilde{{\bf A}}_{+2}^{\rm EQ}]^*$, $\tilde{{\bf \Pi}}_{2,-1}^{(e)}=[-\tilde{{\bf \Pi}}_{2,+1}^{(e)}]^*$, and $\tilde{{\bf \Pi}}_{2,-2}^{(e)}=[-\tilde{{\bf \Pi}}_{2,+2}^{(e)}]^*$.
Hence, once again, apart from the prefactors, both approaches are straightforwardly related to each other, thus confirming the validity of the general expression given in Eq. \eqref{EqS107} up to the electric quadrupole case.

\appendix*\section{Appendix: Angular spectrum representation of the magnetic dipole and its relation with the electric dipole}

From a similar procedure as in the previous sections for the electric dipole and quadrupole, in order to obtain the angular spectrum of an oscillating magnetic dipole, we start by considering its associated vector potential \cite{Jacksonsm}:
\begin{equation}
{\bf A}^{\rm MD}({\bf r})=\frac{i\mu k}{4\pi}\left(1-\frac{1}{ikr}\right)\left({\bf e}_r\times{\bf m}\right)\frac{e^{ikr}}{r},
\label{EqS114}
\end{equation}
where ${\bf m}$ is the \textit{magnetic dipole moment}. The corresponding spectral amplitude is obtained from its partial Fourier transform:
\begin{eqnarray}
\nonumber\tilde{{\bf A}}^{\rm MD}(\kappa_x,\kappa_y;z)&=&\left(\frac{k}{2\pi}\right)^2\iint_{-\infty}^{+\infty}{{\bf A}^{\rm MD}(x,y,z)e^{-ik\left(\kappa_xx+\kappa_yy\right)}dxdy}\\
&=&\frac{i\mu k^3}{16\pi^3}\iint_{-\infty}^{+\infty}{\left(1-\frac{1}{ik\sqrt{x^2+y^2+z^2}}\right)\left({\bf e}_r\times{\bf m}\right)\frac{e^{ik\sqrt{x^2+y^2+z^2}}}{\sqrt{x^2+y^2+z^2}}e^{-ik\left(\kappa_xx+\kappa_yy\right)}dxdy}.
\label{EqS115}
\end{eqnarray}
Making the change of variables to cylindrical coordinates, as given in Eq. \eqref{EqS27}, it follows that
\begin{eqnarray}
\nonumber\tilde{{\bf A}}^{\rm MD}(\kappa_x,\kappa_y;z)&=&\frac{i\mu k^3}{16\pi^3}\int_{0}^{+\infty}{\int_{0}^{2\pi}{\left(1-\frac{1}{ik\sqrt{R^2+z^2}}\right)\left({\bf e}_r\times{\bf m}\right)\frac{e^{ik\sqrt{R^2+z^2}}}{\sqrt{R^2+z^2}}e^{-ik\kappa_R R\cos{(\theta-\phi)}}Rd\theta}dR}\\
&=&\frac{\mu k^3}{8\pi^2}\int_{0}^{+\infty}{\left(1-\frac{1}{ik\sqrt{R^2+z^2}}\right)\boldsymbol{\varOmega}\frac{e^{ik\sqrt{R^2+z^2}}}{R^2+z^2}RdR},
\label{EqS116}
\end{eqnarray}
where
\begin{equation}
\!\!\!\boldsymbol{\varOmega}\equiv\boldsymbol{\varOmega}_{R}+\boldsymbol{\varOmega}_z=RJ_1(k\kappa_RR)\left[m_z\left(\sin{\phi}{\bf e}_x-\cos{\phi}{\bf e}_y\right)\!+\!\left(m_y\cos{\phi}-m_x\sin{\phi}\right){\bf e}_z\right]-izJ_0(k\kappa_RR)\left(m_y{\bf e}_x-m_x{\bf e}_y\right).
\label{EqS117}
\end{equation}
Similarly to the electric quadrupole case, special care must be taken with the integral involving the $z$-dependent contribution. We thus perform the integration separately for the $R$- and the $z$-dependent contributions as follows:
\begin{eqnarray}
\nonumber\tilde{{\bf A}}^{\rm MD}_{R}(\kappa_x,\kappa_y;z)&=&\frac{\mu k^3}{8\pi^2}\int_{0}^{+\infty}{\left(1-\frac{1}{ik\sqrt{R^2+z^2}}\right)\boldsymbol{\varOmega}_{R}\frac{e^{ik\sqrt{R^2+z^2}}}{R^2+z^2}RdR}\\
\nonumber&\xrightarrow[z\to 0]{}&\frac{\mu k^3}{8\pi^2}\int_{0}^{+\infty}{\left(1-\frac{1}{ikR}\right)\left[m_z\left(\sin{\phi}{\bf e}_x-\cos{\phi}{\bf e}_y\right)+\left(m_y\cos{\phi}-m_x\sin{\phi}\right){\bf e}_z\right]J_1(k\kappa_RR)e^{ikR}dR}\\
&=&-\frac{\mu k^2}{8\pi^2}\frac{\kappa_R}{\kappa_z}\left[m_z\left(\sin{\phi}{\bf e}_x-\cos{\phi}{\bf e}_y\right)+\left(m_y\cos{\phi}-m_x\sin{\phi}\right){\bf e}_z\right];
\label{EqS118}
\end{eqnarray}
\begin{eqnarray}
\nonumber\tilde{{\bf A}}^{\rm MD}_{z}(\kappa_x,\kappa_y;z)&=&\frac{\mu k^3}{8\pi^2}\int_{0}^{+\infty}{\left(1-\frac{1}{ik\sqrt{R^2+z^2}}\right)\boldsymbol{\varOmega}_{z}\frac{e^{ik\sqrt{R^2+z^2}}}{R^2+z^2}RdR}\\
\nonumber&=&\frac{i\mu k^3}{8\pi^2}\left(m_y{\bf e}_x-m_x{\bf e}_y\right)\int_{0}^{+\infty}{\left(\frac{1}{ik\sqrt{R^2+z^2}}-1\right)\frac{e^{ik\sqrt{R^2+z^2}}}{R^2+z^2}RzJ_0(k\kappa_RR)dR}\\
\nonumber&=&\left\{\begin{matrix*}[l]
u=J_0(k\kappa_RR)&\quad \Longrightarrow \quad & du=-k\kappa_RJ_1(k\kappa_RR)dR\\
dv=\displaystyle\left(\frac{1}{ik\sqrt{R^2+z^2}}-1\right)\frac{Rz}{R^2+z^2}e^{ik\sqrt{R^2+z^2}}dR&\quad \Longrightarrow \quad & v=\displaystyle\frac{ize^{ik\sqrt{R^2+z^2}}}{k\sqrt{R^2+z^2}}
\end{matrix*}\right\}\\
\nonumber&=&\frac{i\mu k^3}{8\pi^2}\left(m_y{\bf e}_x-m_x{\bf e}_y\right)\left[\left.\frac{izJ_0(k\kappa_RR)e^{ik\sqrt{R^2+z^2}}}{k\sqrt{R^2+z^2}}\right|_{R=0}^{R\to\infty}+\int_{0}^{+\infty}{\frac{iz\kappa_R}{\sqrt{R^2+z^2}}J_1(k\kappa_RR)e^{ik\sqrt{R^2+z^2}}dR}\right]\\
&\xrightarrow[z\to 0]{}&\pm \frac{\mu k^2}{8\pi^2}\left(m_y{\bf e}_x-m_x{\bf e}_y\right).
\label{EqS119}
\end{eqnarray}
Hence, summing up the above results we obtain that
\begin{eqnarray}
\nonumber\tilde{{\bf A}}^{\rm MD}(\kappa_x,\kappa_y;0)&=&\frac{\mu k^2}{8\pi^2}\frac{\kappa_R}{\kappa_z}\left[m_z\left(-\sin{\phi}{\bf e}_x+\cos{\phi}{\bf e}_y\right)-\left(m_y\cos{\phi}-m_x\sin{\phi}\right){\bf e}_z\right]\pm \frac{\mu k^2}{8\pi^2}\left(m_y{\bf e}_x-m_x{\bf e}_y\right)\\
\nonumber&=&\frac{\mu k^2}{8\pi^2}\frac{1}{\kappa_z}\left[\left(\pm m_y\kappa_z-m_z\kappa_y\right){\bf e}_x+\left(m_z\kappa_x\mp m_x\kappa_z\right){\bf e}_y+\left(m_x\kappa_y-m_y\kappa_x\right){\bf e}_z\right]\\
&=&\frac{\mu ck^2}{8\pi^2n}\frac{1}{\kappa_z}\frac{\left[{\bf m}\times{\bf k}^{\pm}\right]}{\omega}.
\label{EqS120}
\end{eqnarray}
Then, the spectral amplitudes of the EM fields for the magnetic dipole reads
\begin{eqnarray}
\tilde{\bf E}^{\rm MD}(\kappa_x,\kappa_y;0)&=&\frac{ic}{kn}[k^2\tilde{\bf A}^{\rm MD}_0-{\bf k}^{\pm}({\bf k}^{\pm}\cdot\tilde{\bf A}^{\rm MD}_0)]=\frac{i\omega k}{8\pi^2}\frac{\mu}{\kappa_z}\left[{\bf m}\times{\bf k}^{\pm}\right],
\label{EqS121}\\
\tilde{\bf H}^{\rm MD}(\kappa_x,\kappa_y;0)&=&\frac{i}{\mu}[{\bf k}^\pm\times \tilde{\bf A}^{\rm MD}_0]=\frac{ik}{8\pi^2}\frac{1}{\kappa_z}\left\{k^2{\bf m}-{\bf k}^{\pm}\left({\bf k}^{\pm}\cdot{\bf m}\right)\right\},
\label{EqS122}
\end{eqnarray}
where $\tilde{\bf A}^{\rm MD}_0\equiv\tilde{\bf A}^{\rm MD}(\kappa_x,\kappa_y;z=0)$. Comparing the above expressions for the magnetic dipole with that given in Eqs. \eqref{EqS40} and \eqref{EqS41} for the electric dipole we find that there exists a relationship between them. Indeed, taking into account the duality principle and the symmetry of Maxwell's equations \cite{Ishimarusm}, we can get the EM fields of the magnetic dipole from that of the electric dipole by simply performing the following transformations:
\begin{equation}
\tilde{{\bf E}}^{\rm ED}\longrightarrow \sqrt{\frac{\mu}{\varepsilon}}\tilde{{\bf H}}^{\rm MD}, \qquad \tilde{{\bf H}}^{\rm ED}\longrightarrow -\sqrt{\frac{\varepsilon}{\mu}}\tilde{{\bf E}}^{\rm MD},\qquad {\bf p}\longrightarrow\frac{n}{c}{\bf m}.
\label{EqS123}
\end{equation}
Notice that, unlike the substitutions pointed out in Ref. \cite{Picardi2017sm}, the above transformation rule does not require the interchange of $\varepsilon$ and $\mu$, thereby becoming more useful for dealing with EM fields in the same medium \cite{Ishimarusm}.



\begin{thebibliography}{90}
\bibitem{footnote_source} Notice that, in the context of guiding systems, the terminology electric and magnetic, referring to the nature of the optical source, is often replaced by transverse magnetic (TM) and transverse electric (TE), respectively.
\bibitem{footnote_limit} It should be noted that the limit $z\to0$ in Eq. \eqref{Eq15} is actually a formal representation that yields a physically consistent meaning to the singular behavior at the origin \cite{Mandel}. In practice, though, this calculation is to be carried out by considering Eq. \eqref{Eq16}, and taking the corresponding limit either on the index $l$ or $m$.
\bibitem{footnote_applet} The notation developed here is very well suited to computationally implement, and we provide a calculator for the angular spectra of arbitrary EM multipoles \cite{link_applet}.
\end{thebibliography}
\begin{thebibliography}{90}
\bibitem{footnote1sm} In the context of guiding systems, the terminology electric and magnetic, referring to the nature of the optical source, is often replaced by transverse magnetic (TM) (or $p$-polarized) and transverse electric (TE) (or $s$-polarized) modes, respectively.
\bibitem{footnote2sm} Notice that the matching between $\mathcal{I}_{l,m}$ and $P_{l'}^{m'}$ works since $\mathcal{I}_{l,m}$ appears multiplied by $P_{l-1}^m$, and $\mathcal{I}_{l,m\pm 1}$ by $P_{l}^m$.
\bibitem{footnote3sm} It should be noted that the limit $z\to0$ in Eq. \eqref{EqS106} is actually a formal representation that yields a physically consistent meaning to the singular behavior at the origin. However, in practice, this calculation is to be carried out by considering Eq. \eqref{EqS105}, and taking the corresponding limit either on the index $l$ or $m$. Notice that, by doing this, one has to assume deliberately that $l$ and/or $m$ are continuous variables.
\end{thebibliography}
\end{document}